\documentclass[twocolumn,showpacs,graphicx,floatfix,eqsecnum,prb]{revtex4-1}
\usepackage{amssymb}
\newcommand{\be}{\begin{equation}}
\newcommand{\ee}{\end{equation}}
\newcommand{\bea}{\begin{eqnarray}}
\newcommand{\eea}{\end{eqnarray}}
\newcommand{\bwt}{\begin{widetext}}
\newcommand{\ewt}{\end{widetext}}
\newcommand{\bk}{{\bf k}}
\newcommand{\bq}{{\bf q}}
\newcommand{\bp}{{\bf p}}
\newcommand{\bkp}{{\bf k}'}

\newcommand{\bv}{{\bf v}}
\newcommand{\ek}{\varepsilon_{\mathbf k}}
\newcommand{\ekp}{\varepsilon_{{\mathbf k}'}}
\newcommand{\ep}{\varepsilon_{\mathbf p}}

\newcommand{\I}{\mathrm{Im}}
\newcommand{\R}{\mathrm{Re}}
\newcommand{\bsu}{\begin{subequations}}
\newcommand{\esu}{\end{subequations}}

\usepackage{amsmath}
\usepackage{epsf}
\usepackage[latin9]{inputenc}
\usepackage{amsmath}
\usepackage{amssymb}
\usepackage{graphicx}
\usepackage{esint}

\begin{document}

\title{First-Matsubara-frequency rule in a Fermi liquid. Part I: Fermionic self-energy}

\author{Andrey V. Chubukov$^1$ and Dmitrii L. Maslov$^2$}

\affiliation{$^1$Department of Physics, University of Wisconsin-Madison, 1150 Univ. Ave., Madison, WI 53706-1390 \\
$^2$Department of Physics, University of Florida, P.O. Box 118440, Gainesville, FL 32611-8440}

\begin{abstract}
We analyze in detail
 the fermionic self-energy
$\Sigma (\omega, T)$  in a Fermi liquid (FL)
at finite temperature $T$ and frequency $\omega$.
We consider
 both canonical FLs -- systems in spatial dimension $D >2$,
 where
  the leading term in the fermionic self-energy is analytic [the retarded $\I\Sigma^R (\omega, T) = C (\omega^2 +\pi^2 T^2)]$, and non-canonical FLs
 in $1<D <2$,
 where
  the leading term in $\I\Sigma^R (\omega, T)$ scales as $T^D$ or $\omega^D$. We relate
 the $\omega^2 + \pi^2 T^2$ form to
 a special property of the self-energy --
 \lq\lq the first-Matsubara-frequency rule\rq\rq\/,
 which stipulates that
 $\Sigma^R(i\pi T,T)$ in
 a canonical FL  contains an
  $\mathcal{O}(T)$
  but
  no $T^2$ term.
   We show that in any $D >1$ the next term after  $\mathcal{O}(T)$ in   $\Sigma^R(i\pi T,T)$ is of order $T^D$
  ($T^3\ln  T$ in $D=3$).
    This $T^D$ term comes from only forward-
   and  backward scattering,
     and
     is  expressed in terms of fully renormalized
     amplitudes for these processes. The overall prefactor of the $T^D$ term
      vanishes in the
       \lq\lq local approximation\rq\rq\/, when the
       interaction can be approximated by its value
       for the initial and final
        fermionic
        states
        right on the Fermi surface.
       The local approximation
      is justified near a Pomeranchuk instability, even if
      the
      vertex corrections
      are
      non-negligible.  We
      show that the strength of the first-Matsubara-frequency rule is amplified in the local approximation,
       where
       it states that not only
       the
       $T^D$ term vanishes
       but
       also that
      $\Sigma^R(i\pi T,T)$
      does not contain any terms
      beyond
      $\mathcal{O}(T)$.
   This rule
     imposes
        two constraints on the
      scaling form of
          the self-energy:
          upon replacing
          $\omega$ by $i\pi T$,
         $\I\Sigma^R (\omega, T)$ must vanish and
            $\R\Sigma^R (\omega, T)$ must reduce to $\mathcal{O}(T)$.
           These two
            constraints should be taken into consideration in
             extracting scaling
            forms of
          $\Sigma^R (\omega, T)$ from experimental and numerical data.
\end{abstract}
\maketitle

\section{introduction}

Properties of single particle and collective excitations
in strongly interacting electron systems
 continue to attract
  substantial
  interest
of the condensed-matter community.
This interest is stimulated by
the avalanche of discoveries of new materials, many of which fall into a category of strongly correlated electron systems, and by
 advances in experimental techniques, which allow
 one to extract, with
 good accuracy, the
 single-particle self-energy from angle-resolved photoemission (ARPES) data
 and the two-particle or
 \lq\lq optical\rq\rq\/ self-energy
from the real and imaginary parts of the optical conductivity.

  One of the most
  actively explored directions in the study of strongly correlated electron systems is a search for
   non-Fermi liquids
   (non-FLs) -- systems in which electrons interact so strongly that
  they completely lose coherence.
 Many newly discovered systems
 were classified as non-FLs  because their electron spectral functions, extracted from ARPES, are quite broad. However, a broad spectral function is an indication, but not the proof,
that the system in question is
 a non-FL, as the Landau criterion for the FL
  only requires that the spectral function must be sharp for fermions in the immediate vicinity of the Fermi surface (FS).
  A mathematical formulation
   of this requirement is that
   the imaginary part of the retarded self-energy
   $\I\Sigma^R(\omega)$ must be much smaller than $\omega + \R\Sigma^R(\omega)$ at the smallest $\omega$.
  This does not preclude that at higher frequencies $\I \Sigma^R (\omega)$ can become
   comparable to $\omega + \R\Sigma^R(\omega)$ or even exceed it.

 To satisfy  the Landau criterion, $\I \Sigma^R (\omega)$  has to scale as $\omega^{1+a}$ with $a >0$.
  The original argument by Landau, based on the Pauli principle and the assumption of analyticity, yields $\I \Sigma^R (\omega) \propto \omega^2$, i.e., $a=1$.
  Microscopic calculations show
 that
 $\I\Sigma^R (\omega)$
  does indeed scale as $\omega^2$ in a 3D FL. The same holds for all
   \lq\lq fractional\rq\rq\/ dimensions $D>2$.
  For
  $D \leq 2$, the analyticity is, however, broken:
 $\I\Sigma^R (\omega)$ scales as $\omega^2\ln|\omega|$ in $D=2$
  and as
 $|\omega|^D$ in $D<2$.  Still, by  Landau criterion, these systems are FLs, as long as $D >1$. Hereafter we refer to
  systems in which $\I \Sigma^R (\omega) \propto \omega^2 $ as
 \lq\lq  canonical FLs\rq\rq\/ , and
  to
   systems in which
  $\I \Sigma^R (\omega) \propto \omega^{1+a}$ with $0<a<1
  $ as
 \lq\lq non-canonical FLs\rq\rq\/.

 The goal of this paper is to analyze the form of the self-energy in
 both conventional and non-conventional FLs
   at finite frequency $\omega$ and temperature $T$.
  We will be particularly interested
  in
   how general is a certain property of the self-energy, which we will be referring to as the
    \lq\lq first-Matsubara-frequency rule\rq\rq\/ or, for brevity, as the  \lq\lq first-Matsubara rule\rq\rq\/. This rule
  states that the self-energy
  $\Sigma (\omega_m, T)$,
  evaluated at
  discrete
  Matsubara points
   $\omega_m=\pi T(2m+1)$,
    exhibits a special behavior
  at the first fermionic Matsubara frequency,
 $\omega_{0} = \pi T$,
  namely, $\Sigma (\pi T, T)$  does not contain terms higher than
 $T$.
 (The same happens at $\omega_{-1} = - \pi T$.)
    This rule was
   proven in the past for particular cases of the electron-phonon \cite{fowler65,engelsberg78}
 and screened Coulomb\cite{martin03,adamov}  interactions.
  In the former case, this rule is sometimes being referred  to as a \lq\lq Fowler-Prange theorem\rq\rq\/.
 \cite{wasserman96}

Although the first-Matsubara rule operates on the imaginary frequency axis, it is relevant to properties of physical fermions with real frequencies:
  it requires that the retarded self-energy
  $\Sigma^R (\omega, T)$, with $\omega$ replaced by $i \pi T$, should not contain terms beyond ${\mathcal O}(T)$, and thus
    imposes a constraint on the interplay between the $\omega$ and $T$ terms in  $\Sigma^R (\omega, T)$.

A 3D FL provides a simple example of how the first-Matsubara rule works.
 To order
 $\omega^2$ and $T^2$, we have in this case
    $\mathrm{Re} \Sigma^R (\omega, T) = \lambda \omega$, with no $\omega T$ term,  and  $\mathrm{Im} \Sigma^R (\omega, T) =C(\omega^2 + \pi^2 T^2)$, with  a factor of exactly $\pi^2$
  in front of $T^2$.
 At $\omega = i \pi T$,  $\mathrm{Im} \Sigma^R$
 vanishes and $\mathrm{Re} \Sigma^R$
 becomes
 of order $T$, hence the total $\Sigma^R (i\pi T, T)$ contains only an ${\mathcal O}(T)$ term but no $T^2$ term.

In this paper, we analyze the validity of the first-Matsubara rule beyond the
conventional FL paradigm.
The proof of this rule in prior work\cite{fowler65,engelsberg78,martin03,adamov} was based
 on demonstrating the  nullification of the  leading term in imaginary part of the self-energy at the first Matsubara frequency.
 We show here that the first-Matsubara rule does not hold beyond
 the leading order for
 conventional FLs, and does not hold at all for unconventional FLs.
 Our primary finding is that  $\Sigma^R(i\pi T,T)$ scales as $T^D$ in all $D$
 (with an extra $\ln T$ factor in $D=3$); however, the consequences of this finding
 are different for conventional and unconventional FLs.
  For conventional FLs, i.e.,  for $2<D <3$,
 the  $T^D$ term is still subleading to $T^2$, and thus
  the first-Matsubara rule holds to order $T^2$.
    For unconventional FLs, i.e., for $1<D <2$,
    the $T^D$ term is of the same order as the leading terms in $\I\Sigma^R (\omega, T)$,
    and thus the
    the first-Matsubara rule
    is violated.
    In $D =2$, which is a marginal case between conventional and unconventional FLs,
    $\mathrm{Im} \Sigma^R(\omega, T) \propto (\omega^2 + \pi^2 T^2) \ln |\omega|
    + \mathcal{O}(\omega^2, T^2)$.
    While
    the logarithmic term vanishes at $\omega = i\pi T$,
    the
    $T^2
    $
    term does not.  As a result,
 the
    first-Matsubara rule is satisfied to logarithmic accuracy
     but not beyond.

 We find that  for
  $1<D<3$,
     the $T^D$ term in $\Sigma^R (i\pi T, T)$ is
        universal,  i.e.,
     independent of the upper cutoff of the theory.
      Furthermore, its prefactor is
     expressed via exact spin and charge components of the forward-
     and backscattering amplitudes.

At the same time, we find
that the first-Matsubara rule holds to {\em all} orders in $T$  in both conventional and non-conventional FLs,
 if the
 effective interaction between fermions, which includes dynamic
  screening by particle-hole bubbles,
 is assumed to connect only the states right on the Fermi surface.
    Hereafter we refer to this approximation as the \lq\lq local approximation\rq\rq\/, as it is generally valid when
   bosons which mediate interaction between fermions are slow compared to fermions.\cite{acs,chubukov_locality,georges:96}

 We show that, within the local approximation,
   the first-Matsubara rule relies only on the analytic properties of the local susceptibility.
   For the electron-phonon interaction, this approximation is a key
     ingredient of the Eliashberg theory,~\cite{eliash62:se} and the small parameter which controls this approximation
      is the ratio of the Debye frequency to Fermi energy.
   We consider here the case of an electron-electron interaction. In certain limits it can be approximated by an effective interaction mediated by collective modes of fermions in  the spin or charge channel.  The collective modes are generally not slow compared to fermions themselves (their velocity is of order
   of the
   Fermi velocity),
   but they
   do
   become slow near a Pomeranchuk instability, when the correlation length for critical collective modes diverges.
    As a result of this divergence,
     the system
     generates a low-energy scale,
      below which near-critical collective modes become overdamped and slow down.\cite{chubukov_locality,rech}
    The local approximation for collective modes is a
      necessary but not sufficient condition
      for
      the Eliashberg theory,
      as the latter also requires vertex corrections to be small. In the case of collective modes,
      vertex corrections are not controlled by the same parameter which makes the local approximation valid,~\cite{aim,rech,cm_05,sslee,metl_sachdev,mc_10} and are not necessary small.~\cite{sslee,metl_sachdev,metl_sachdev_2,senthil,senthil_1}
 We show that the smallness of vertex corrections is not required for the first-Matsubara rule to work -- the local approximation is sufficient.

     We analyze the local approximation in more detail and
      show that the first-Matsubara rule
      imposes two conditions:
      1)
      $\mathrm{Im} \Sigma^R (i \pi T, T)$ vanishes
    to all orders in $T$,
    and 2)
    $\mathrm{Re} \Sigma^R (i \pi T, T)$ contains an
    ${\mathcal O}(T)$
    term
    but all higher order terms in $T$ vanish.
 These
       two conditions
        are non-trivial
       because, beyond the conventional FL paradigm,
  $\Sigma^R (\omega, T)$ {\em cannot} be obtained from the $T=0$ result by a simple replacement
   $\omega\to \sqrt{\omega^2+\pi^2T^2}$. This is true for the subleading $\omega^3, T^3$ terms in a 3D FL, and
   also
   for the leading $\omega^D$, $T^D$  terms in
    non-conventional FLs.
   In particular, $\mathrm{Im} \Sigma^R (\omega, T)$  in
    non-conventional FLs has a complex form which is very different from
    $(\omega^2 + \pi^2 T^2)^{D/2}$, and  $\mathrm{Re} \Sigma^R (\omega, T)$
    also contains a complex dependence on  $\omega$ and $T$ at order $\omega^D$, in addition to the $\lambda \omega$ term. Nevertheless, as long as the local approximation is
    applicable,
    $\mathrm{Im} \Sigma^R(\omega, T)$ vanishes at $\omega =i\pi T$,
    and $\mathrm{Re} \Sigma^R
     (i\pi T, T)$ reduces to $i \pi \lambda T$.

Finally, we show that the first-Matsubara rule holds within the local approximation even for a non-FL, e.g., for a system in $D \leq 3$ right at a
  Pomeranchuk instability,
   except that in this case the coefficient $\lambda$ in $\Sigma^R(i \pi T,T) = i \pi T \lambda$  diverges  as $T \to 0$.  In particular, the first-Matsubara rule holds for a marginal FL and for an itinerant 2D system at a nematic
   quantum critical point (QCP).

The rest of the paper is organized as follows. In Sec.~\ref{sec:2}, we review the derivation of
 the single-particle self-energy to order $T^2$ and $\omega^2$  in a conventional FL,  and show where the relation between the $\omega^2$ and $T^2$ terms comes
  from.
In Sec.~\ref{sec:5}, we discuss the self-energy outside of the conventional FL paradigm. We show that, in general,
   the self-energy contains terms of order $T^D$,  which do not vanish when $\omega$ is replaced by $i\pi T$.
   The case of $D =2$ is marginal, and we consider it separately.
In Sec.~\ref{sec:4}, we discuss the
 self-energy within the local approximation. We show that, at order $T^D$,  $\Sigma^R(\omega, T)$
 has quite a complex dependence on the ratio $\omega/T$, yet the prefactor of the $T^D$ term vanishes
 at
 $\omega=i\pi T$.
 We consider in detail 2D and 3D FL's,  a 2D system at a nematic QCP, and also a marginal FL.
    We discuss under what conditions the local approximation is valid
    in all these cases.
    We also discuss in this Section
    how
   one should properly
   construct
    the self-energy along real frequency axis to
  make sure that a replacement of $\omega$ by $i \pi T$ agrees with the analytical continuation of the self-energy into the upper
  half-plane.
  We present our conclusions in Sec~\ref{sec:concl}.

In the subsequent paper,~\cite{MCII} we discuss the constraints imposed by the first-Matsubara rule
on the $\Omega/T$ scaling the optical conductivity $\sigma (\Omega, T)$ of a FL,
 and the consequences of these constraints
 for the experiment.

Throughout the paper, we
 denote
 the
 retarded self-energy along
 the
 real frequency axis as $\Sigma^R_{\bk} (\omega, T)$
and
 the
 self-energy along the Matsubara axis as
$\Sigma_{\bk}(\omega_m, T)$,
 where ${\bk}$ is the electron (quasi)momentum.
  We set the overall sign of the
 retarded
 self-energy via
 \be
  G^{R}_k (\omega,T) =
\frac{1}{ \omega + \Sigma^R_{\bk} (\omega, T) - \varepsilon_{\bk}},
 \ee
 where $\ek$ is the electron dispersion,
  and
  define the Matsubara self-energy in such a way that it is real on the Fermi surface,
  i.e.,
  \be
 G_{\bk}  (\omega_m,T) =
 \frac{1}{i
 \left[\omega_m + \Sigma_{\bk} (\omega_m, T)
 \right] - \varepsilon_{\bk}}.
 \label{def}
   \ee

 \section{Single-particle self-energy: canonical Fermi liquid}
\label{sec:2}

In this Section, we briefly review
 the derivation of the
  scaling forms for the self-energy in a conventional FL to order  $\omega^2$ and $T^2$:
 $\I\Sigma^R(\omega,T) \propto \omega^{2}+\pi^2 T^{2}, ~ \R\Sigma^R(\omega,T) = \lambda \omega$
 with no
 $\mathcal{O}( \omega T)$ terms in either of these quantities.
 We first show how these forms are obtained in the perturbation theory,
 then
 use the Eliashberg's argument\cite{eliash62:se} to
 generalize the derivation to an arbitrary order in the interaction, and
 finally  relate these forms
  to the first-Matsubara rule -- a special property of the self-energy
  at the first fermionic Matsubara frequency
   $\omega_{m=0,-1} = \pm
    \pi T$ (Sec.~\ref{sec:FPT}).

\subsection{Perturbation theory}
\label{sec:2a}

We consider a system of fermions on a lattice with
 single-particle
dispersion
  $\varepsilon _{{\bf k}}$.
  We assume that the Fermi surface does not have nested parts and is away from the van Hove singularities
  but otherwise arbitrary.
  Near the FS, $\ek$ can be
approximated as $\varepsilon _{{\bf k}}={\bf v}_{{\bf k}_{F}}\cdot \left( \bk-\bk_{F}\right)$, where ${\bf k}_{F}$ is a
vector pointing in the direction of ${\bf k}$  and residing on the FS, and ${\bf v}_\bk=\boldsymbol{\nabla}
_{{\bf k}}\varepsilon _{{\bf k}}$.
We will see that $\omega^2$ and $T^2$ terms in $\I\Sigma^R_{\bk}
(\omega,T)$ come from low-energies
 where the linear approximation is valid. Having this in mind, we follow a conventional reasoning of a FL theory,
 set the upper
 cutoff of the theory with the linearized
  dispersion at some energy $\Lambda$ (generally
  comparable to the bandwidth, $W$), and absorb all renormalizations from energies between $\Lambda$ and $W$  into non-singular renormalizations of the effective mass and quasiparticle residue $Z$.
  The bare Green's function of low energy fermions is then given by
  \be
  G^R_{{\bf k}}\left(\omega\right)
=Z_{{\bf k}_{F}}/\left( \omega-\varepsilon _{{\bf k}} + i \delta\right),
  \label{la_1}
  \ee
  where $\delta
  >0 $ is infinitesimally small and $Z_{{\bf k}_{F}}$, in general, varies along the FS.
We further assume that fermion-fermion interaction, $U_{\bq}$,
is static and
 non-singular for all $
 {\bf  q}$ connecting points on the FS, including $q=0$. This is the case for, e.g.,
 a
 screened  Coulomb interaction.

\begin{figure}[t]
\includegraphics[width=0.5\textwidth]{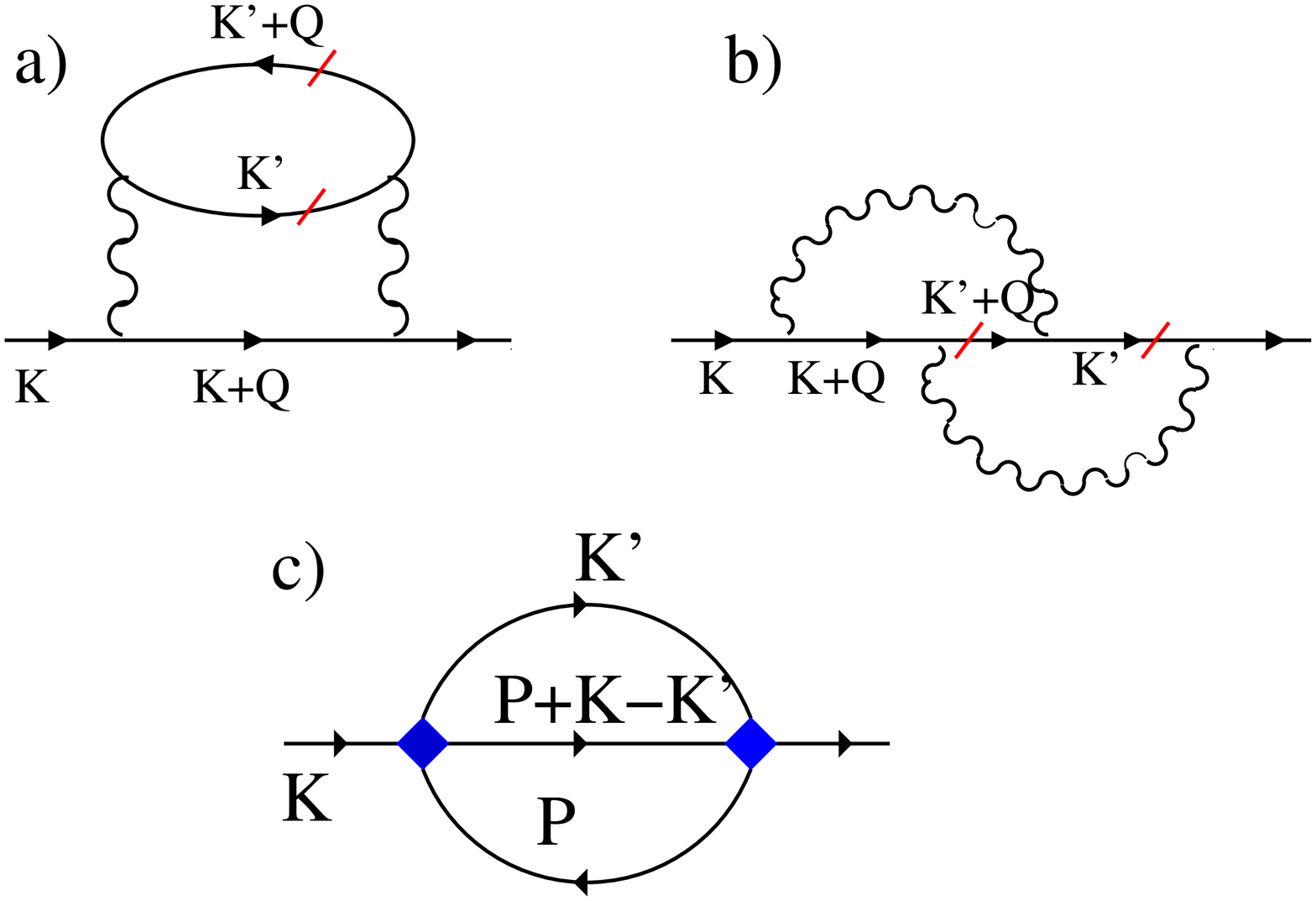}
\caption{(color on-line). Diagrams for the self-energy. $K=(\omega_m,\bk)$, $K'=(\omega_{m'},\bkp)$, $Q=(\Omega_n,\bq)$.}
\label{fig:selfenergy}
\end{figure}

 The lowest-order diagrams which contribute to the imaginary part of the fermionic self-energy are shown in Fig.~\ref{fig:selfenergy}.
 The imaginary part of the fermionic self-energy arises from the convolutions of two Green's functions
marked by slanted dashes in  Fig.~\ref{fig:selfenergy}. In diagram $a$, such a
convolution is just a particle-hole bubble, the imaginary part of which
scales linearly with the bosonic frequency $\Omega$. In diagram $b$,
this convolution involves the momentum-dependent
interaction,  but the result still scales
linearly with $\Omega$.

 To see this in more detail, we write down a Matsubara
form of the self-energy from diagram $a)$ and obtain  $\Sigma^R
 {\bf k}
(\omega,T)$ by analytic continuation.
 With our definition for
 the
 self-energy (\ref{def}), we have
\begin{equation}
\Sigma^a_{{\bf k}}\left( \omega _{m},T\right) =
-i
T\sum_{\Omega _{n}}\int_{{\bf q}}U({\bf q})^{2}G_{{\bf k+q}}\left( \omega _{m}+\Omega _{n}\right)
\Pi _{{\bf q}}\left( \Omega _{n}\right) ,  \label{sigma_a_M}
\end{equation}
where
\begin{equation}
\Pi _{{\bf q}}\left( \Omega _{n}\right) = 2T\sum_{\omega _{m^{\prime }}}\int_{{\bf k}^{\prime
}}G_{{\bf k}^{\prime }}\left( \omega
_{m^{\prime }}\right) G_{{\bf k}^{\prime }+{\bf q}}\left( \omega _{m^{\prime
}}+\Omega _{n}\right)   \label{Pi_M}
\end{equation}
with $\Omega _{n}=2\pi nT$ and $\int_{{\bf l}}\equiv \int d^{D}{\bf l/}%
\left( 2\pi \right) ^{D}$.  Performing analytic continuation in both (\ref
{sigma_a_M}) and (\ref{Pi_M}),
 we obtain  the retarded self-energy along the real frequency axis
\begin{subequations}
\bwt
\begin{eqnarray}
\Sigma _{{\bf k}}^{R,a}\left( \omega ,T\right)  &=&\int_{{\bf q}}U _{{\bf q}%
}^{2}\int \frac{d\Omega }{2\pi }\left[ \coth \frac{\Omega }{2T}G_{{\bf k+q}%
}^{R}\left( \omega +\Omega \right)
\mathop{\rm Im}
\Pi _{{\bf q}}^{R}\left( \Omega \right) +\tanh \frac{\Omega +\omega }{2T}%
\mathop{\rm Im}
G_{{\bf k+q}}^{R}\left( \omega +\Omega \right) \Pi _{{\bf q}}^{A}\left(
\Omega \right) \right]  \label{sigma_a1}\\
\Pi _{{\bf q}}^{R}\left( \Omega \right)  &=&2\int_{{\bf k}^{\prime }}\int
\frac{d\omega ^{\prime }}{2\pi }\!\!\left[ \tanh \frac{\omega ^{\prime }}{2T}%
\mathop{\rm Im}
G_{{\bf k}^{\prime }}^{R}\left( \omega ^{\prime }\right) G_{{\bf k}^{\prime
}+{\bf q}}^{R}\left( \omega ^{\prime }+\Omega \right) +\tanh \frac{\omega
^{\prime }+\Omega }{2T}G_{{\bf k}^{\prime }}^{A}\left( \omega ^{\prime
}\right)
\mathop{\rm Im}%
G_{{\bf k}^{\prime }+{\bf q}}^{R}\left( \omega ^{\prime }+\Omega \right) %
\right].
\label{Pi}
\end{eqnarray}
\ewt
\end{subequations}
Extracting the imaginary parts of Eqs.~(\ref{sigma_a1}) and (\ref{Pi}) and using the relations $\tanh(x/2)=1-2n_F(x)$ and $\coth(x/2)=2n_B(x)+1$, where $n_F(x)$ and $n_B(x)$ are the Fermi and Bose functions, correspondingly, we obtain
\bsu
\bea
\I\Sigma^{R,a}_\bk(\omega,T)&=&\int_\bq U^2_{\bq}\int\frac{d\Omega}{\pi}\left[n_B(\Omega)+n_F(\omega+\Omega)\right]\notag\\
&&\times\I G^R_{\bk+\bq}(\omega+\Omega)\I\Pi^R_\bq(\Omega)\label{imsigma}\\
\I\Pi^R_\bq(\Omega)&=&2\int_{\bkp}\int\frac{d\omega'}{\pi}\left[n_F(\omega'+\Omega)-n_F(\omega')\right]\notag\\
&&\times\I G^R_{\bkp}(\omega')\I G^R_{\bkp+\bq}(\omega'+\Omega).\label{imp}
\eea
\esu
Equation (\ref{imp}) can be-rewritten as
\bwt
\bea
\I\Pi^R_\bq(\Omega)&=&\int_{\bkp}\int d\omega'\left[n_F(\omega'+\Omega)-n_F(\omega')\right]\oint \frac{dA_{\bkp_F}}{v_{\bkp_F}(2\pi)^{2}} Z_{\bkp_F}Z_{\bkp_F+\bq}\int d\ekp
\delta\left(\omega'-\ekp\right)\delta\left(\omega'+\Omega-\epsilon_{\bkp+\bq}\right)
\label{imp_1}
\eea
\ewt
where $dA_{\bkp_F}$ is
 the
 element of the $
D-1
 $-dimensional FS. The integral over $\ekp$ gives
$\delta(\omega'+\Omega-\epsilon_{\bkp+\bq})\left\vert_{\ekp=\omega'}\right.$. The role of this $\delta$-function
is to impose a constraint on the angle between $\bkp$ and $\bq$. Since this angle is not, in general, small, it suffices to resolve this constraint at $\omega'=\Omega=0$  because, as subsequent integration will show, $\omega'\sim\Omega\sim \max\{\omega,T\}$. The $\delta$-function thus reduces to $\delta(\epsilon_{\bkp_F+\bq})\left\vert_{\ekp=0}\right.$, which means that both the initial and final states are on the FS.
[Notice that this approximation corresponds to expanding the $\delta$-functions in $\max\{\omega,T\}/E_F$ rather than in $\omega/\ek$.]
The integral over $\omega'$ now gives $\int^\infty_{-\infty} d\omega' \left[n_F(\omega')-n_F(\omega'+\Omega)\right]=\Omega$, and $\I\Pi^R_\bq(\Omega)$ reduces to $\Omega$ multiplied by a function of $\bq$, averaged over the FS:
\bea
\I\Pi^R_\bq(\Omega)&=&-\frac{\Omega}{(2\pi)^{2}}\oint \frac{dA_{\bkp_F}}{v_{\bkp_F}} Z_{\bkp_F}Z_{\bkp_F+\bq}\delta(\epsilon_{\bkp_F+\bq})\left\vert_{\ekp=0}\right.\notag\\
\eea
For small $q$, the prefactor of $\Omega$ behaves as $1/q$. Substituting $\I\Pi^R_\bq(\Omega) \propto \Omega$ into (\ref{imsigma}),
 and applying the same procedure
  as above to integrate over
  the
  momentum, we obtain, for $
  \bk=\bk_F
  $
 \be
\I \Sigma^{R,a}_{\bk_F} (\omega,T) = 2 C_a  \int^\infty_{-\infty} d\Omega\Omega\left[n_B(\Omega)+n_F(\omega+\Omega)\right]
 \label{la_4}
 \ee
 with
\bea
C_a&=&\frac{\pi}{2(2\pi)^{D-1}}\int_\bq\oint \frac{dA_{\bkp_F}}{v_{\bkp_F}}Z_{\bk_F+\bq} Z_{\bkp_F}Z_{\bkp_F+\bq}\notag\\&&\times\delta(\varepsilon_{\bk_F+\bq})\delta(\epsilon_{\bkp_F+\bq})U_{\bq}^2.\label{ca}
\eea
 [A factor of 2 in (\ref{la_4}) is introduced for future convenience.]
 The frequency integral in Eq.~(\ref{la_4}) is readily evaluated
\be
\int^\infty_{-\infty} d\Omega\Omega\left[n_B(\Omega)+n_F(\omega+\Omega)\right]=\frac{1}{2}\left(\omega^2+\pi^2 T^2\right).
\label{sigma_int}\ee
hence
\be
\I \Sigma^{R,a}_\bk(\omega,T)=C_a\left(\omega^2+\pi^2 T^2\right),
\label{10}
\ee

 We can now specify what actually makes the analysis above
  applicable only to conventional FLs rather than to all FLs: it is an assumption that the integral in Eq.~(\ref{ca}) is convergent in the infrared. Power counting shows that the integrand behaves as $1/q^2$ for $q\to 0$;
   the integral over $d^{D-1} q$ then
   converges for $D>2$ and diverges for $D\leq 2$. Infrared divergence for
    $D\leq 2$ will modify the $\omega$ and $T$ dependencies of $\Sigma^R
    {\bk_F}
    (\omega,T)$ compared to the canonical form valid
       for $D>2$.

Diagram $b$ is analyzed in a similar way with the only difference that the quantity $U_q\I\Pi_\bq^R
 (\Omega)$ in Eq.~(\ref{imsigma}) is replaced by
\bea
\I{\cal P}^R_{\bq,\bk}(\Omega)&=&\int_{\bkp}\int\frac{d\omega'}{\pi}\left[n_F(\omega'+\Omega)-n_F(\epsilon)\right]U_{\bk-\bkp}\notag\\
&&\times\I G^R_{\bkp}(\epsilon)\I G^R_{\bkp+\bq}(\epsilon+\Omega).\label{imp1}
\eea
Still, $\I{\cal P}^R_{\bq,\bk}(\Omega)$ scales as $\Omega$ for $\Omega\to 0$. Evaluating the integrals in the same way as above, we find
\be
\I \Sigma^{R,b}_{\bk_F}(\omega,T)=C_b\left(\omega^2+\pi^2 T^2\right),
\label{10_1}
\ee
where
\bea
C_b&=&-\frac{\pi}{4(2\pi)^{D-1}}\int_\bq\oint \frac{dA_{\bkp_F}}{v_{\bkp_F}}Z_{\bk_F+\bq} Z_{\bkp_F}Z_{\bkp_F+\bq}\notag\\&&\times\delta(\varepsilon_{\bk_F+\bq})\delta(\epsilon_{\bkp_F+\bq})U_{\bq}U_{\bkp_F -\bk_F}.
\label{ca1}
\eea
 As before, the integral in Eq.~(\ref{ca1}) is convergent for $D>2$.
Comparing $\I \Sigma^{R,a}_{\bk_F}(\omega,T)$ and $\I \Sigma^{R,b}_{\bk_F}(\omega,T)$, we see that they both have the same
 scaling form $\omega^2 + \pi^2 T^2$ and differ only in prefactors which, in general case, depend on $\bk_F$, i.e.,
 on position along the FS.

 The real part of the self-energy can
 be
  obtained either directly, e.g., from Eq.~(\ref{sigma_a1}) for diagram $a$, or via a Kramers-Kronig (KK) transformation of  $\mathrm{Im}\Sigma_\bk
  ^R
   (\omega,T)$
\be
\mathrm{Re} \Sigma^R_{\bk_F} (\omega,T) = \frac{2\omega}{\pi} {\mathcal P} \int_0^\infty d\omega' \frac{\mathrm{Im} \Sigma^R_{\bk_F} ( \omega',T)}{\omega'^2 - \omega^2}
\label{11},
\ee
where ${\mathcal P}$ stands for the principal part.
 The integral is ultraviolet divergent if Eqs.~(\ref{10}) or (\ref{10_1})
is used for  $ \mathrm{Im}\Sigma^R_{\bk_F} (\omega,T)$,
 which implies that,
 to get the correct  form of $\mathrm{Re}\Sigma _{\bk_F} (\omega,T)$ from the KK transformation, one has to use the full form of $ \mathrm{Im}\Sigma^R_{\bk_F} (\omega,T)$ rather than its  low-energy approximation. Nevertheless,
 one can easily make sure that, to quadratic order, $ \mathrm{Re}\Sigma_{\bk_F}(\omega,T) = \lambda_{\bk_F} \omega$ (where $\lambda_{\bk_F}$ varies, in general, along the FS) with no $\omega T$ term.

A  comment
  is in order here.
  By applying (\ref{11})
 to (\ref{10}) or (\ref{10_1}),
 we can only show that there is no "universal", cutoff-independent $
 \omega T$ term in $\mathrm{Re}\Sigma^R_\bk (\omega,T)$, and hence no $T^2$ term in $\mathrm{Re}\Sigma^R_\bk (i\pi T,T)$.  There is still a possibility that a $T^2$ term in $\mathrm{Re}\Sigma^R_\bk (i\pi T,T)$  may come
  from internal frequencies in (\ref{11}) comparable to
  the upper cutoff of the low-energy theory.  We show later, in Sec. \ref{sec:FPT}, that this is not the case, and that only a $T^3$ term emerges from high energies.

\subsection{Arbitrary order in the interaction}
\label{sec:2b}
We now follow the argument by
 Eliashberg\cite{eliash62:se}
who
 showed that the $\omega^2 + \pi^2 T^2$ form  of the self-energy at  finite $T$
  holds to all orders
 in the
  interaction
  (a similar reasoning was also employed by Luttinger \cite{luttinger61} to show that $\I\Sigma^R
  {\bk_F}
  (\omega,T=0)\propto \omega^2$).

The argument is as follows.
In the second-order diagrams,
the $\omega^2 + \pi^2 T^2$
 form
 comes from
 the
  region
  where
 all three intermediate fermions are located within the window of width of order $\omega$ or $T$ near the FS.
 Accordingly, the
 interactions  $U_{\bf q}$ can be approximated by their values
 evaluated for the case when when the initial and final states are on the FS,
 i.e., ${\bf q}={\bf l}_F-{\bf l}'_F$.

 In a self-energy diagram of any order, one can select a cross-section with three low-energy fermions, and sum
 over all other fermions
  without
  assuming that they are near the FS.  The diagrams of this kind can be cast in the form of  Fig.~\ref{fig:selfenergy}c).
   The three selected fermions are near the FS and the shaded
   squares are
   the
   full vertex functions.  Because
    integration over
    the
    fermionic lines already gives
    a function
    quadratic in $\omega$ or $T$,
     one can set $T=0$ in the remainder of the diagram and
     project
      all four external momenta
      onto
       the FS. As long as the full vertex functions do not diverge, they do not affect
        integration over dispersions and frequencies of intermediate fermions.
    Self-energy corrections to fermionic lines are also irrelevant because
    the
     dressed Green's function still has the form of Eq.~(\ref{la_1}) at the lowest energies -- adding one-loop self-energy to Eq.~(\ref{la_1})
    simply
    replaces $i\delta$ by $i C (\omega^2 + \pi^2 T^2)$,
 which has an
 extra power of energy compared to $\omega$ and hence does not affect
 the
 $\omega^2$ and $T^2$ terms in the full self-energy.  As a result,
the
    $\omega^2 + \pi^2 T^2$ form survives to an arbitrary order in the interaction -- self-energy and vertex renormalizations  only affect the overall factor in $\I\Sigma^R_{\bk_F} (\omega, T)$.
  We then have for a conventional FL and to order $\omega^2,T^2$
      \be
     \I\Sigma^R_{\bk_F}(\omega,T) = C \left(\omega^2 + \pi^2 T^2\right).
     \label{la_2}
     \ee
     The prefactor $C$ depends on model parameters, including the cutoff $\Lambda$, and
      is thus non-universal. Substituting this form into
     KK formula, Eq.~(\ref{11}), and using the same arguments as in previous
     section, we find
       \be
     \R\Sigma^R_{\bk_F}(\omega,T) = \lambda \omega \left(1 + 0
     \times
     T \right)
     \label{la_3}
     \ee
     (we spelled  out the $0\times T$ combination
     to emphasize that there $\R
     \Sigma^R(\omega,T)$ does not contain
     an
     $\omega T$ term.)
 The prefactor $\lambda$ is again non-universal.

\subsection{Self-energy along the imaginary  axis:\\
the first-Matsubara-frequency rule}
\label{sec:FPT}
  We now
 show that the scaling form of $\mathrm{Im}\Sigma^R_{\bk_F}(\omega,T)$ in (\ref{la_2}) and the absence of
 the
 $\omega T$ term in
  $\mathrm{Re}\Sigma^R_{\bk_F}(\omega,T)$ are related
   to  a particular behavior of the self-energy
   at
   the first fermionic Matsubara frequency
  $\omega_{m=0} =
  \pi T$ (the same
   behavior holds
   at
    $\omega_{m=-1} = -
    \pi T$).

\subsubsection{Analytic continuation}

Let us first analytically continue  $\mathrm{Im} \Sigma^R_\bk(\omega,T)$ and $\mathrm{Re} \Sigma^R_\bk(\omega,T)$ in Eqs.~(\ref{la_2}) and (\ref{la_3})
into the upper-half  plane of the complex  variable $\omega \to z = z' + iz''$.
 Because $\mathrm{Im} \Sigma^R_\bk(\omega,T)$ and $\mathrm{Re} \Sigma^R_\bk(\omega,T)$ are analytic, their analytic continuation
  reduces to just a  replacement of $\omega$ by $z$.
  The complex function  $\mathrm{Im} \Sigma^R(z,T)\propto z^2+\pi^2T^2$ then vanishes
  at
   $z = i \pi T$
  and
  $
  \R \Sigma^R(z,T)$ becomes $
  i\pi \lambda T$,
  so that the full self-energy
  reduces to $i \pi T \lambda$ and does not contain a $T^2$ term.  This is only true, however,
    if $z$ is replaced by first Matsubara frequency. For any other $\omega_m \neq
    \pi T$, the
    function $\Sigma^R (i\omega_m,  T)$
contains a $T^2$ term.
As we said in the Introduction, we
will refer to this property as to the first-Matsubara-frequency rule (first-Matsubara rule).

It is worth stressing that the rule formulated above applies only to the {\em first} Matsubara frequency.
 This may not seem to be the case if we
 replace $\omega$
 by
  $ i \pi (2m+1)T $
  with arbitrary $m$
 in
$\I\Sigma^R_{\bk_F}(\omega,T)$
 given by
 Eq.~(\ref{imsigma}),
 {\it before} integrating over the bosonic frequency $\Omega$.
 Doing so,
   and using the identity $ n_F (\Omega+i\omega_m)=-n_B (\Omega)$, we seemingly
   find that $\I\Sigma^R(i\omega_m,T)$  vanishes
 not only at $\omega=
  i\pi T$ but also at {\em any} Matsubara frequency $i\pi T(2m+1)$.

  \begin{figure}[t]
\includegraphics[width=0.5\textwidth]{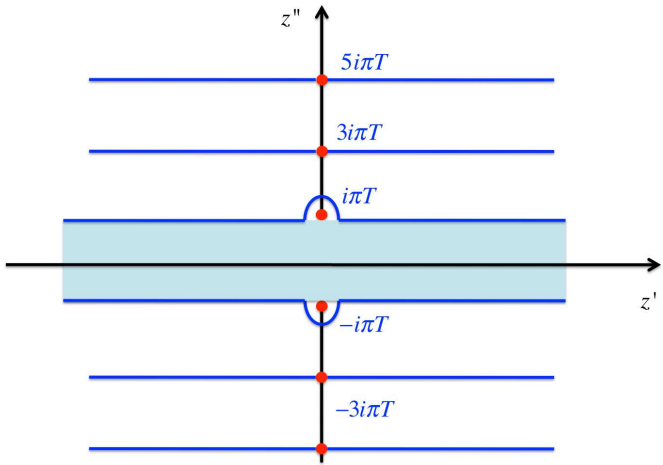}
\caption{(color on-line). Analytic structure of the function $\I\Sigma^R_{\bk_F}(z,T)$ in the complex $z$ plane. Analytic continuation from the real axis is possible  to any point within the shaded region, including the points $\pm i\pi T$, but not beyond this region.}
\label{fig:bcuts}
\end{figure}

 This
 result
 is, however, false because the complex function,
 obtained by analytic continuation of,
 e.g.,
  Eq.~(\ref{imsigma}),
  into the complex plane
   {\em before} the integral over $\Omega$ is performed,
   contains a sequence of branch cuts that run parallel to the real axis and intersect the imaginary axis at the Matsubara frequencies (see Fig.~\ref{fig:bcuts}).
   As a result, the imaginary part of the function
 \be
 F(z)=\int d\Omega \Omega \left[n_B(\Omega)+n_F(\Omega+z)\right].
 \ee
 changes discontinuously at $z=z'+i\pi (2m+1) T$. For example, a discontinuity of $\mathrm{Im}F(z)$ at $z = z' + i \pi T$ is
  \bea
&&\left[ \mathrm{Im}F\left(z'+iT(\pi +\delta/2)\right)-\mathrm{Im}F\left(z'+iT(\pi -\delta/2)\right)\right ]|_{\delta\to 0}\notag\\
 &&=-\delta \int d\Omega \frac{\Omega}
 {\sinh^2\frac{\Omega+z'}{2T}+\delta^2}\approx  2\pi T z'. \eea
 This implies that the substitution $\omega \to z = z' + i z''$ into the integral form of  $\mathrm{Im}\Sigma^R
 {\bk_F}
 (\omega, T)$,
 Eq. ~\ref{la_4}), gives the same
 result for
 $\mathrm{Im}\Sigma^R_{\bk_F}(z, T)$ as the actual analytical continuation
only in the
 region bounded by two branch cuts at $z = z' + i\pi T$ and $z = z' - i \pi T$,  but not
 outside this region.
In other words, the substitution $\omega = i\omega_m$ into (\ref{la_4}) gives the correct result for only for the first,
but not for all Matsubara frequencies.

\subsubsection{Direct proof of the first-Matsubara frequency rule}

The first-Matsubara rule can be also proven directly, by computing the self-energy
 for a
conventional FL in  Matsubara
 frequencies.
 For the electron-electron interaction, this was done in Refs.~\onlinecite{martin03} and \onlinecite{adamov};
  however, the proofs presented in these two papers are valid under two additional assumptions, namely,
 of small-angle scattering and
 of a
 quadratic
 dispersion, $\ek=(k^2-k_F^2)/2m^*$,
 where $m^*$ is the renormalized effective mass.
  In fact, neither of these two assumptions are necessary. In what follows we first consider the case of arbitrary-angle scattering but still keep an assumption of a
  quadratic dispersion, and then generalize
 the argument for an arbitrary dispersion.\\

  \paragraph{{\bf Quadratic dispersion.}}

 To be specific,
 we consider the 3D case; other dimensions can be considered in a similar way.
  The clamshell self-energy diagram (diagram $c$ in Fig.~\ref{fig:selfenergy}) reads
 \bwt
 \bea
&&\Sigma_{{\bk}_F} (\omega_m, T)
 = -i
 Z^3
 T\sum_{\Omega_n} T \sum \omega_{m'} \int \frac{d^3k'}{(2\pi)^3}\int \frac{d^3p}{(2\pi)^3} \nonumber \\
&& \times G_{\bk'}(\omega_m+\Omega_n)G_{\bp}(\omega_{m'})G_{\bp+\bk-\bk'}(\omega_{m'}+\Omega_n)  \Gamma_{\bk_F,\bp_F;\bk'_F,\bp_F+\bk_F-\bk'_F}
\Gamma_{\bk'_F,\bp_F+\bk_F-\bk'_F;\bk_F,\bp_F},
\label{la_15}
\eea
\ewt
where $\Gamma_{\bk,\bp;\bk',\bp'}$ is the renormalized vertex (a filled diamond in Fig.~\ref{fig:selfenergy}c).
Since we have already assumed that the dispersion is isotropic, the $Z$ factor is assumed to be isotropic as well.
 The momentum transfers
   can be arbitrary, but all three intermediate momenta are
   assumed to be
    near the FS; this assumption has already been used in Eq.~(\ref{la_15}).

 To evaluate the momentum integrals, the dispersion $\epsilon_{\bp+\bk-\bk'}$ needs to be expanded in $\ep$, $\ek$, and $\ekp$. For a
 quadratic
 dispersion,  we obtain
  after some algebra
\bea
&&\epsilon_{\bp+\bk-\bk'}=\frac{2k_F^2}{m^*}\sin\frac{\theta_{\bk,\bk'}}{2}\left(\sin\frac{\theta_{\bk,\bk'}}{2}+\cos\theta_{\bp,\bk-\bk'}\right) \notag\\
&&+\ep\left(1+2\sin\frac{\theta_{\bk,\bk'}}{2}\cos\theta_{\bp,\bk-\bk'}\right)\notag\\
&&+\left(\ek+\ekp\right)\left(2\sin^2\frac{\theta_{\bk,\bk'}}{2}+\sin\frac{\theta_{\bk,\bk'}}{2}\cos\theta_{\bp,\bk-\bk'}\right),\notag\\
\label{la_16}
\eea
where $\theta_{{\bf l},{\bf m}}$ is the angle between
 momenta
${\bf l}$ and ${\bf m}$.
A similar analysis can be carried out for any isotropic but not necessarily quadratic dispersion.
 For small-angle scattering $\theta_{\bk,\bk'}\ll 1$, Eq.~(\ref{la_16})  reduces to a familiar form $\epsilon_{\bp+\bq}=\epsilon_{\bp}+v_Fq\cos\theta_{\bp,\bq}$ with $q=2k_F\sin(\theta_{\bk,\bk'}/2)$.
 For
 the momentum $\bp+\bk-\bk'$ to be on the FS, the first term in Eq.~(\ref{la_16}) must be small;
 for generic values of $\theta_{\bk,\bk'}$, this condition amounts to a geometric constraint
\be
\cos\theta_{\bp,\bk-\bk'}=-\sin\frac{\theta_{\bk,\bk'}}{2}
\label{la_17}
\ee
or $\theta_{\bp,\bk-\bk'}=
\pm(\pi+\theta_{\bk,\bk'})/2$. We expand the first term in (\ref{la_16}) around this value as $\theta_{\bp,\bk-\bk'}=\pm(\pi+\theta_{\bk,\bk'})/2-\alpha$ with $\alpha\ll 1$, and set $\alpha=0$ in the remaining two terms.
  This gives
\bea
&&\epsilon_{\bp+\bk-\bk'}=v^*_Fk_F\alpha\sin\theta_{\bk,\bk'}\notag\\
&&+\ep\cos^2\theta_{\bk,\bk'}+(\ek+\ekp)\sin^2\frac{\theta_{\bk,\bk'}}{2},
\label{la_18}
\eea
 where $v_F^*=k_F/m^*$.
 Substituting the last result into (\ref{la_16}), we obtain
\bwt
\bea
&&\Sigma_{{\bk}_F} (\omega_m,T)=
-2 i
Z^3
T\sum_{\Omega_n} T \sum_{\omega_{m'}} \left(\frac{k_F^2}{v_F}\right)^2\int \frac{d\ekp}{(2\pi)^2}\int
d\theta_{\bk,\bkp}\sin\theta_{\bk,\bk'}\cos\frac{\theta_{\bk,\bkp}}{2}\int \frac{d\ep}{(2\pi)^2}\int d\alpha \frac{1}{i(\Omega_n+\omega_m)-\ekp}
\frac{1}{i\omega_{m'}-\ep}\notag\\
&&
\times
\frac{1}{i(\omega_{m'}+\Omega_n)-v_Fk_F\alpha\sin\theta_{\bk,\bk'}-\ep\cos^2\theta_{\bk,\bk'}-(\ek+\ekp)\sin^2\frac{\theta_{\bk,\bk'}}{2}}
\Gamma_{\bk_F,\bp_F;\bk_F',\bp_F+\bk_F-\bk_F'}\Gamma_{\bk_F',\bp_F+\bk_F-\bk_F';\bk_F,\bp_F}.\notag\\
\label{la_18_1}
\eea
\ewt
 Constraint (\ref{la_17}) is assumed
 to be imposed on the momenta entering both vertices in the last equation.
 The integral over $\alpha$ gives
\be
-i\frac{\pi}{v^*_F k_F\sin\theta_{\bk,\bk'}}\mathrm{sgn}(\omega_{m'} +\Omega_n),
\ee
 while the integral over $\ep$ gives  $-i\pi\mathrm{sgn}\omega_{m'}$. Summing the product of the two sign functions over $\omega_{m'}$, we obtain a \lq\lq local\rq\rq\/, i.e., integrated over the momentum, polarization bubble as a sum of two terms: $-|\Omega_n|/\pi$ and a constant, proportional to the ultraviolet cutoff of the theory.
  The constant contributes only to the $\mathcal{O}(T)$ term in $\Sigma
 _ {\bk_F}
   (\pi T,T)$, and we consider it separately later.
  The $|\Omega_n|$ term is the one relevant to our purposes as we need to verify that
   $\Sigma_{\bk_F} (\pi T,T)$ does not contain a $T^2$ contribution.  The prefactor of the $|\Omega_n|$  term is given by
   \bea
&&2C=\left( \frac{
m^*Z
}{2\pi}\right)^3\int d\theta_{\bk,\bk'}\cos\frac{\theta_{\bk,\bkp}}{2}\Gamma_{\bk_F,\bp_F;\bk_F',\bp_F+\bk_F-\bk_F'}\notag\\
&&\times\Gamma_{\bk_F',\bp_F+\bk_F-\bk_F';\bk_F,\bp_F}.
\eea
 The remaining integral over $\ekp$ gives $\mathrm{sgn}(\omega_m+\Omega_n)$,
   and the self-energy becomes
\bea
&&\Sigma_{\bk_F}(\omega_m,T)=
 -2
  C \pi T\sum_{\Omega_n}\mathrm{sgn}(\omega_m+\Omega_n)|\Omega_n| +\dots \nonumber \\
&&
\label{ms}\eea
where dots stand for
$\mathcal{O}(T)$ terms.
Summation over $\Omega_n$ is straightforward, and we obtain
 \bea
\Sigma_{\bk_F}(\omega_m,T)=
C \left(\pi^2 T^2  - \omega^2_m\right) +\dots
\label{ms_1}
\eea
 The $T^2$ term in the Matsubara self-energy
 obviously vanishes for $\omega_m=\pm
  \pi T$.

\paragraph{{\bf Arbitrary dispersion.}}

Equation (\ref{ms_1}) is also valid for an arbitrary fermionic dispersion,
with the only difference that  the prefactor $C$ now depends on the position
on the FS.
To see how this works, we expand $\epsilon_{\bp+\bk-\bkp}$ near
 a FS.  Knowing that $\ek$ and $\ekp$ drop out anyway, we set them to zero and expand $\epsilon_{\bp+\bk-\bkp}$ around $\bp_F$ as
\be
\varepsilon_{\bp+\bq_F}=\varepsilon_{\bp-\bp_F+\bp_F+\bq_F}=\epsilon_{\bp_F+\bq_F}+\ep\frac{v^{||}_{\bp_F+\bq_F}}{v^{||}_{\bp_F}},
\label{lt1}
\ee
where $\bq_F\equiv \bk_F-\bk'_F$, $v^{||}_{\bf l}\equiv \bv_{\bf l}\cdot{\hat \bp}_F$,
 ${\hat \bp}_F\equiv\bp_F/p_F$,
 and we suppressed $^*$ in $v_F$ for brevity.

Substituting this expansion into (\ref{la_15})
 and replacing  integrals over 3D momenta by integrals over the FS and over the electron energy, we obtain
\bwt
\bea
\Sigma_{{\bk}_F} (\omega_m,T)&=&
-i
T\sum_{\Omega_n} T \sum_{\omega^{'}_n}
 \oint\frac{dA_{\bk'_F}}{(2\pi)^3v_{\bk'_F}}\oint \frac{dA_{\bp_F}}{(2\pi)^3v_{\bp_F}}
 {\tilde Z}^3
 \int d\ekp\int d\ep \frac{1}{i(\Omega_n+\omega_m)-\ekp}
\frac{1}{i\omega^{'}_n-\ep}\notag\\
&&
\times
\frac{1}{i(\omega^{'}_n+\Omega_n)-\varepsilon_{\bp_F+\bq_F}-\ep\frac{v^{||}_{\bp_F+\bq_F}}{v^{||}_{\bp_F}}}
\Gamma_{\bk_F,\bp_F;\bk_F',\bp_F+\bq_F}\Gamma_{\bk_F',\bp_F+\bq_F;\bk_F,\bp_F},
\label{u_1}
\eea
\ewt
where ${\tilde Z}^3\equiv Z_{\bk'_F}Z_{\bp'_F}Z_{\bp_F+\bk_F-\bk'_F}$.
The condition that all three internal fermions are located near the FS implies that,
 at fixed $\bq_F$, the angle between $\bp_F$ and $\bq_F$ must be such that the first term in
 $\varepsilon_{\bp_F+\bq_F}$
 is small. Suppose that, at fixed $\bq_F$, the constraint $\varepsilon_{\bp^{0,i}_F+\bq_F}=0$ is satisfied for a set of symmetry-related points on the FS,
 $\bp_F^{0,i}$.  The vector $\bp_F$ spans a narrow solid angle around each of  $\bp_F^{0,i}$; therefore we can expand the dispersion as
 $\varepsilon_{\bp_F+\bq_F}=(\bp_F-\bp_F^{0,i})\cdot{\tilde \bv}$, where ${\tilde \bv}\equiv {\bv}_{\bp^{0,i}+\bq_F}$.
 Since $\bq_F$ is still fixed,  it can be chosen as the polar axis of a local spherical system, in which a point on the FS is described by an equation $p_F=r(\theta,\phi)$.  Vectors $\bp_F^{0,i}$ are parametrized as $p_F=r(\theta^{0,i},\phi^{0,i})$; correspondingly,  $p_F=r(\theta^{0,i}-\alpha,\phi^{0,i}-\beta)\approx p_F^{0,i}-\alpha r_{\theta}-\beta r_{\phi}$ where $r_{\theta}$
and $r_{\phi}$ are the partial derivatives of $r$ with respect to $\theta$ and $\phi$, respectively, evaluated at the point
$(\theta^{0,i},\phi^{0,i})$.  Suppose that  ${\tilde \bv}$ makes angle $\gamma$ with the polar axis and, without a loss of generality, assume that $x$ axis
 belongs to the plane formed by vectors ${\tilde \bv}$ and $\bq_F$.
Then, $\cos\theta_{\bp_F^{0,i},{\tilde\bv}}=\cos\theta^{0,i}\cos\gamma$ and $\cos\theta_{\bp_F^{0,i},{\tilde\bv}}\approx
\left(\cos\theta^{0,i}+\alpha\sin\theta^{0,i}\right)\cos\gamma$ to linear order in $\alpha$. Substituting all of the results above into Eq.~(\ref{lt1}),
we obtain
\be
\varepsilon_{\bp+\bq_F}=\alpha\left(p_F^{0,i}\sin\theta^{0,i}-r_\theta\cos\theta^{0,i}\right)\cos\gamma+\ep\frac{v^{||}_{\bp^{0,i}_F+\bq_F}}{v^{||}_{\bp^{0,i}_F}},
\label{lt2}
\ee
which generalizes Eq.~(\ref{la_18}) for the arbitrary dispersion case.
 The measure of integration over the area $dA_{\bp_F}$ reduces to
\be
\frac{dA_{\bp_F}}{v_{\bp_F}}\approx \frac{(p_F^{0,i})^2}{v^{||}_{\bp^{0,i}_F}}\sin\theta^{0,i} d\alpha d\beta.\ee

The rest of the calculations proceeds in the same way as for the
 quadratic-dispersion case;
 namely,  integrating first over $\alpha$, then  $\ep$ and, finally, over $\ek$, we reproduce the same product of the three sign factors as before.  The final expression for the self-energy
 reduces to that in Eq. (\ref{ms}) with a different prefactor, which varies over the FS.

 Interestingly, we found that the seemingly obvious result that
 frequency summation in Eq.~(\ref{ms}) yields (\ref{ms_1})
   can be reproduced only with a considerable effort if one uses the Euler-Maclaurin  formula to sum over $\Omega_
   n
   $. Namely,
  one has to keep not only the \lq\lq conventional\rq\rq\/ terms
  with the integral over $
  n$ and derivatives of the summand at $
  n=0$, but also the \lq\lq remainder\rq\rq\/ term which which is often neglected when the Euler-Maclaurin formula is applied
  in practice.
    We discuss this issue in Appendix~\ref{app_a}.\\

\subsubsection{
  The linear-in-$T$ term
 in the Matsubara self-energy}

Finally, we
consider in more detail the $O(T)$ contribution to $\Sigma_{\bk_F} (\pi T,T)$.  For definiteness, we focus on the 3D case
and restrict to
 quadratic
 dispersion. If we integrate
 in Eq.~(\ref{la_15})
 over
 $\ep$
  and $\ekp$ in infinite limits, as we did earlier in this Section, and
  retain
 a constant term
 [denoted as $\Pi (0)$)]
  instead of the $|\Omega|$ term  in
   the local polarization bubble, we obtain
 \be
\Sigma_{\bk_F}(\pi T,T)= \lambda
T  \sum_{\Omega_n}\mathrm{sgn}(
\pi T
+\Omega_n)
\label{ms_2}
\ee
where $\lambda
 \propto
 \Pi(0)$.
Because
 only the $n=0$ term contributes to the sum,
$\Sigma_{\bk_F}(\pi T,T) =  \lambda T$.

This result holds
 only
 if we integrate over
$\ep$
 and
$\ekp$
 in  infinite limits.
  Since, however,
  we have set the cutoff of our low-energy theory
 to
 $\Lambda$,
 integrations over $\ep$ and $\ekp$ should, strictly speaking, be
 performed between $-\Lambda$ and $\Lambda$. The magnitude of
 $\lambda$ then depends on the ratio $\Lambda/E_F$ and reduces to the previous result only
  for $\Lambda \gg E_F$.
 In the
  opposite
   limit
   of
   $\Lambda\ll E_F$, which is more appropriate for systems
   in
   which  $E_F$ is of the same order as the bandwidth,  $\lambda$ is much smaller,
   namely, $\lambda\sim(\Lambda/E_F)\ln(E_F/\Lambda)$.
 We show this in Appendix~\ref{app_new}.
We also checked
 if there is
 a  $T^2$ contribution to $\Sigma_{\bk_F}(\pi T,T)$ at
 finite $\Lambda$
 but
 found
 no
such term. The next term
after the $\lambda T$
  is of order $T^3/\Lambda^2$.  This one is irrelevant to our purposes, as later in the text we show that in a generic 3D FL there are universal terms of order $(T^3/E_F^2) \ln{E_F/T}$, which are parametrically larger than a non-universal $T^3$ term.

\section{
Single-particle self-energy: \\
 non-canonical Fermi liquids and higher-order terms in canonical Fermi liquids}
\label{sec:5}
We remind the reader that
 the analysis in the previous
 Section
 relied
  on the
 assumption that the momentum integrals,
 incorporated into the prefactor $C$
  in
   Eq.~(\ref{la_2}) for $\I\Sigma^R_{\bk_F} (\omega, T)$,
 are free from singularities. These
 integrals include quasiparticle renormalization factors,
 the
 effective interaction between the quasiparticles, and the prefactor
 of the $\Omega$ term in
 the imaginary part of
  the polarization operator [see Eqs.~(\ref{ca}) and (\ref{ca1})].
  The quasiparticle
 renormalization factors  and
 the effective interaction are non-singular at small $q$, but the prefactor
  of the $\Omega$ term scales as $1/q$ and may give rise to infra-red divergencies.
 The momentum integral
 in the expression for $C$
  is over
  the
  $D-1$ components
 of ${\bq}$
 lying in a plane tangential to the $D$-dimensional FS.
  This integral
  converges
  for $D >2$,
  i.e., in  a conventional FL,
  but diverges
  for $D <2$,
  i.e.,
 in a non-conventional FL.

 The issue
  we discuss in this
 Section
 is whether the first-Matsubara rule holds  in a non-conventional FL,
  and  in a conventional FL
  beyond the $T^2$ order.
   We will show in this Section that
   the next after the $\mathcal{O}(T)$ term in $\Sigma_{\bk_F}(\pi T,T)$ scales as $T^D$ for any $D$, i.e.,
$\Sigma_{\bk_F}(\pi T,T)=\lambda T+dT^D$ .

     The $T^D$ term is
     subleading to the
     $T^2$
     one in a conventional FL ($D>2$),
     and thus the first-Matsubara rule holds to order $T^2$ in this case.
     However, the leading terms in a non-conventional FL ($1<D \leq 2$)
     are also of the $T^D$ order, and thus the first-Matsubara rule does not hold in this case.
   In the next Section, we show that the first-Matsubara rule holds to all orders in $T$
   for any $D$
   near QCP, when
   the local approximation becomes valid.

 We consider first the marginal case of $D =2$,
 and then discuss the cases of $2<D<3$, $D =3$, and  $1<D<2$.

 \subsection{$D=2$}
\label{sec:d=2}

   In $D=2$, the self-energy is non-analytic:
    $ \mathrm{Im} \Sigma^R_{\bk_F}(\omega,T) \propto \omega^2 \ln |\omega|$ at $T=0$ and $T^2 \ln T$ at $\omega =0$,
    \cite{hodges,chaplik,guiliani}
     while the first subleading term in $ \mathrm{Re} \Sigma^R_{\bk_F} (\omega,T)$ scales as $\omega|\omega|$ at $T=0$ and as $T^2 \mathrm{sgn} \omega$
    for
    $\omega\ll T$.\cite{cm_05}
 To logarithmic
 accuracy,
 the scaling form of  $\mathrm{Im} \Sigma_{\bk_F}^R (\omega,T)$ is given by
 \cite{martin03,adamov,dassarma,laikhtman,macdonald,reizer,narozhny,cm_03,vignale}
   \be
  \mathrm{Im} \Sigma_{\bk_F}^R (\omega,T) =  C_2 \left(\omega^2 + \pi^2 T^2\right) \ln{\frac{
  \Lambda}{|\omega|}},
  \label{35}
  \ee
  where $C_2$
  is a constant.
  By the KK relation,
   \be
  \mathrm{Re} \Sigma_{\bk_F}^R (\omega,T) = \lambda \omega  -\frac{\pi C_2}{2} \mathrm{sgn} \omega \left(\omega^2 + \pi^2 T^2\right).
  \label{35_1}
  \ee
  At this level, the first-Matsubara rule is obviously satisfied.
 Beyond logarithmic accuracy, however, the situation is different, as we will now see.

 Let us first  calculate the self-energy in Matsubara frequencies. Consider diagram $a)$ in Fig.~\ref{fig:selfenergy}.
 The corresponding formula for the self-energy is given by  Eq.~(\ref{sigma_a_M}).
  We explore an earlier observation~\cite{cm_05,cm_03,vignale,efetov_1}
 that  the non-analytic contributions to the fermionic self-energy come
  from forward-
 and
  backscattering
  rather than
  from
 scattering
 by an arbitrary
 angle.
  The internal structures of diagrams
  with
  forward scattering and backscattering are the same, i.e., it is sufficient to analyze only one
    of these
    two
    contributions. We consider forward scattering, i.e., focus on small momentum transfers $q$,
    and also assume that the FS is isotropic (a circle). Consequently, the self-energy does not depend
    on the position on the FS but we will still keep the subscript $\bk_F$ which indicates that the self-energy is evaluated on the FS,
    as opposed to the self-energy evaluated away from the FS also considered in this Section.

 At small $q$, the  polarization bubble behaves as
   \be
   \Pi_{\bq} (\Omega_n) = -
  \frac{m}
  {\pi} \left(1 -  \frac{| \Omega_n|}{\sqrt{\Omega^2_n  + (v_F q)^2}}\right).
  \label{26}
  \ee
  The constant term in $\Pi$ gives rise to an ${\mathcal O}(T)$ term in $\Sigma_{\bk_F }(i \pi T,T)$. We neglect it
  for
  now
  but will
   re-instate
  it  in the final
   result for
   $\Sigma_{\bk_F} (\pi T,T)$.
 Keeping the
 dynamic
 part in (\ref{26}) and introducing  polar coordinates for momentum integration,
  we obtain for the forward-scattering contribution
  of diagram $a)$
  to the self-energy
  at arbitrary momentum $\bk$
 \bea
&&\Sigma _{\bk}(\omega_m,T) =
-i A_2 T \sum_{ \Omega_n}  \int \frac{q d q d
 \phi}{(2\pi)^2}  \nonumber \\
 &&\times  \frac{1}{i(\omega_m +  \Omega_n)
 -\ek-
  v_F q \cos
  \phi} \frac{| \Omega_n|}{\sqrt{\Omega^2_n + (v_F q)^2}},
\label{27}
\eea
where $A_2 =
4\pi u^2(0)/m$
  and $u(0)\equiv mU_{\bq=0}/2\pi$ is the dimensionless coupling constant for forward scattering.
Integrating over $\theta$, we obtain
 \bea
&&\Sigma _{\bk}(\omega_m,T) =
-A_2 T \sum_{ \Omega_n}  \int \frac{d qq}{2\pi}  \nonumber \\
 &&\times  \frac{\mathrm{sgn}(\omega_m+\Omega_n)}{ \sqrt{\left(\omega_m+\Omega_n+i\ek\right)^2+(v_F q)^2}} \frac{| \Omega_n|}{\sqrt{\Omega^2_n + (v_F q)^2}}.
\label{27_dm}
\eea
 First, we discuss the self-energy on the FS. Substituting $\ek=0$ into Eq.~(\ref{27_dm})
and
  integrating over $q$ up to $\Lambda/v_F$,  we obtain
\bea
&&\Sigma_{\bk_F} (\omega_m,T) = -\frac{
 T A_2}{2\pi v^2_F} \sum_{ \Omega_n} | \Omega_n| \mathrm{sgn} (\omega_m+  \Omega_n) \nonumber\\
&&\times\ln\left[\frac{\sqrt{\Lambda^2 + \Omega^2_m} +\sqrt{\Lambda^2 + (\omega_m + \Omega_n)^2}}{| \Omega_n| + |\omega_m +  \Omega_n|}\right].
\label{29}
\eea
For $\omega_m = \pi T$,
 the last result
 reduces to
 \begin{widetext}
\be
\Sigma_{\bk_F} (\pi T,T) =-\frac{
 T^2 A_2}{v^2_F} \sum_{n=1}^\infty m \ln\left[\frac{2n-1/2}{2n+1/2}~ \frac{
\left({\bar \Lambda}^2 + n^2\right)^{1/2}
 +
 \left({\bar \Lambda}^2 + (n +1/2)^2
 \right)^{1/2}}
 {\left({\bar \Lambda}^2 + n^2\right)^{1/2}
 +
 \left({\bar \Lambda}^2 + (n -1/2)^2
 \right)^{1/2}}
 \right],
\label{30}
\ee
\end{widetext}
where ${\bar \Lambda} = \Lambda/(2\pi T) \gg 1$.  To evaluate the frequency sum, we notice that the second
fraction
 under the logarithm is
 close to
 unity
 in both regions of $m$ that are relevant for the sum, namely,
 for $n \ll {\bar \Lambda}$ and for $m \sim {\bar \Lambda}$, when
 $n\ll  n^2$.  In either case,
 \be
\ln  \frac{
\left({\bar \Lambda}^2 + n^2\right)^{1/2}
 +
 \left({\bar \Lambda}^2 + (n +1/2)^2
 \right)^{1/2}}
 {\left({\bar \Lambda}^2 + n^2\right)^{1/2}
 +
 \left({\bar \Lambda}^2 + (n -1/2)^2
 \right)^{1/2}}
\approx \frac{n}{2({\bar \Lambda}^2 +n^2)}.
\label{31}
\ee
With this simplification, the sum over $n$ can be evaluated exactly.
Performing summation, and adding the ${\mathcal O}(T)$ contribution from the static part of the polarization bubble,
  we obtain
\be
\Sigma_{\bk_F} (\pi T,T) =
\pi T \lambda - \frac{A_2 T^2}{2\pi v^2_F} \left(K + \frac{\pi\ln 2}{4}\right)
\label{32}
\ee
where $ \lambda \sim (A_2 \Lambda)/v^2_F$ is a non-universal constant and $K =  0.9160$ is the Catalan's constant  ($K + \pi\ln{2}/4 = 1.460$).
We see that $\Sigma (\pi T,T)$  does contain
 a universal, i.e., cutoff-independent,
$T^2$ term.
 We recall that there is no such term in $D >2$, when the self-energy is analytic to order $T^2$.
The presence of such a term
 in $D=2$ implies that the first-Matsubara rule breaks down once the self-energy becomes non-analytic.

    For completeness, we also reproduced Eq.~(\ref{32}) by evaluating first $\mathrm{Im}\Sigma^R_{\bk_F} (\omega,T)$
   and then  evaluating $\Sigma_{\bk_F} (\pi T,T)$ using the general KK relation between
the Matsubara self-energy and $ \mathrm{Im} \Sigma^R_{\bk_F}(\omega,T)$
   \be
\Sigma_{\bk_F} (\omega_m,T) = \frac{2
 \omega_m}{\pi} \int_0^\infty d \omega \frac{\mathrm{Im} \Sigma^R_{\bk_F}(\omega,T)}{\omega^2 +\omega_m^2}
\label{33}
\ee
 Applying spectral representation to Eq.~(\ref{27}) and integrating over the momentum, we obtain for $\omega >0$
 \bea
 && \mathrm{Im} \Sigma^R_{\bk_F}(\omega,T) =\frac{A_2}{4\pi^2 v^2_F}
  \int^{\Lambda-\omega}_{-\Lambda} d
  \Omega\Omega\left[n_B (\Omega) + n_F (\Omega+\omega)\right] \nonumber \\
 && \times \ln{\frac{\left[\sqrt{\Lambda^2 -\Omega^2} + \sqrt{\Lambda^2 - (\Omega+\omega)^2}\right]^2}{\omega |\omega +2\Omega|}},
 \label{34}
 \eea
 if $\omega<2\Lambda$, and  $\mathrm{Im} \Sigma^R_{\bk_F}(\omega,T)=0$ otherwise.
 To logarithmic accuracy, this expression reduces to Eq.~(\ref{35}).

  Substituting (\ref{34}) into (\ref{33}) and
  setting
  $
  \omega_m=\pi T
  $, we find that the
  main
  logarithmic term in $\mathrm{Im} \Sigma^R_{\bk_F} (\omega,T)$
  contributes
  only
  to the $\mathcal{O}(T)$ term in  $\Sigma_{\bk_F} (
   \pi T,T)$.  The violation of the first-Matsubara rule comes from
   the subleading $\mathcal{O}(\omega^2)$  and $\mathcal{O}(T^2)$ terms.
    We obtained the first term in (\ref{32}) analytically and reproduced
     the second term by integrating  over $\omega$ in (\ref{33})  numerically.

 A complete expression for $\Sigma_{\bk_F}(\pi T,T)$  to second order in the interaction
 contains contributions from
 diagrams $a)$ and $b)$ in  Fig.~\ref{fig:selfenergy}.
  Each of these diagrams
  contains contributions
from the  interaction
 with momentum transfers equal to zero
and to $2k_F$
with amplitudes
$U(0)$ and $U(2k_F)$,
 correspondingly.
  Collecting all these contributions, we obtain a complete result for $\Sigma_{\bk_F}(\pi T,T)$   to second order in the interaction as
 \cite{cm_05,vignale}
\bea
&&\Sigma_{\bk_F} (\pi T,T) =
 \pi T \lambda \nonumber \\
&&-
 \frac{T^2}{
  2E_F} \left[3u^2(0) + 2 u^2(
  2k_F) - 2 u(0) u(
  2k_F
  )\right] \left(K + \frac{\pi\ln 2}{4}\right),\nonumber \\
\label{la_la}
\eea
where
$u(q) = m U_{\bq}/(2\pi)$.
  The combination of the
  coupling constants in (\ref{la_la}) can be
  expressed via the spin and charge components of the forward ($f$)
   and backscattering ($b$) amplitudes, $\Gamma^f$ and $\Gamma^b$,
   defined by
\bea
 \Gamma^{f,b}_{\alpha\gamma;\beta\delta} = \Gamma^{f,b}_c \delta_{\alpha \beta} \delta_{\gamma\delta} +
 \Gamma^{f,b}_s  {\boldsymbol \sigma}_{\alpha \beta}\cdot {\boldsymbol \sigma}_{\gamma\delta},
\label{la_7}
\eea
where subscripts $c$ and $s$ stand for \lq\lq charge\rq\rq\/ and \lq\lq spin\rq\rq, respectively.
To first order in $U_{\bq}$,
\be
\Gamma^f_c = - \Gamma^f_s = u(0),~\Gamma^b_c= 2 u(0) - u(
2k_F), ~\Gamma^b_s = - u(
2k_F).
\label{la_8}
\ee
Using these relations,
 one can re-express Eq.~(\ref{la_la}) as
\bea
&&\Sigma _{\bk_F}(\pi T, T) =
 \pi T \lambda -
  \frac{T^2}{
8 E_F}  \left(K + \frac{\pi\ln 2}{4}\right)
\nonumber \\
&&\times  \left[2 \left\{(\Gamma^b_c)^2 + 3 (\Gamma^b_s)^2\right\} + (\Gamma^f_c)^2 + 3 (\Gamma^f_s)^2\right].
\label{32_2}
\eea

Equation (\ref{32_2}) can be extended to a FL with an arbitrary interaction.
  One can show, using the same arguments as in Refs.~\onlinecite{cm_05,cmm}, that the
  self-energy
  still contains
  the same combination of
  forward- and backscattering amplitudes,
  except for in a general case
 $\Gamma^b_{c,s}$ and $\Gamma^f_{c,s}$ are expressed not via $u(0)$ and $u(
 2k_F)$ but rather via fully renormalized four-fermion
 vertices
  $\Gamma (\bk,\bk;\bk,\bk), ~\Gamma(\bk,-\bk;\bk,-\bk)$, and $\Gamma(\bk,-\bk,-\bk,\bk)$, which may depend on both transferred and total momenta.
Explicitly,
we have
 \bea
\Gamma^f_c &=& \frac{Z^2 m^*}{2\pi}\Gamma (\bk,\bk;\bk,\bk); ~ \Gamma^f_s  -\frac{Z^2m^*}{2\pi}\Gamma (\bk,\bk;\bk,\bk) \nonumber \\
\Gamma^b_c &=& \frac{Z^2 m^*}{2\pi}\left[2\Gamma (\bk,\bk,-\bk;\bk,-\bk) - \Gamma (\bk,-\bk;-\bk,\bk) \right], \nonumber\\
~ \Gamma^b_s &=& -\frac{Z^2 m^*}{2\pi}\Gamma (\bk,-\bk;-\bk,\bk).
\label{la_8_1}
\eea
A complete
expression for the self-energy at the first Matsubara frequency is
\bea
&&\Sigma_{\bk_F} (\pi T, T) =
 \pi T \lambda -
 \frac{T^2}{
 8 E_F}  \left(K + \frac{\pi\ln 2}{4}\right) \frac{m^*}{m Z}
\nonumber \\
&&\times  \left[2 \left\{(\Gamma^b_c)^2 + 3 (\Gamma^b_s)^2\right\} + (\Gamma^f_c)^2 + 3 (\Gamma^f_s)^2\right].
\label{32_2_1}
\eea
 There is one additional complication:
the result in Eq.~(\ref{32_2_1}) is actually based on the expansion of the polarization bubble in frequency:
for free fermions, this amounts to replacing (\ref{26}) by $\Pi_\bq(\Omega_n)=-(m/\pi)\left(1-|\Omega_n|/v_Fq\right)$.
The static part of $\Pi_\bq(\Omega_n)$
 produces the $T$ term in (\ref{32_2_1}),
 while the
  (smaller) dynamic part
  produces the $T^2$ term.
 At weak coupling,  one can safely set the lower limit of integration over $q$ to
 zero,
 because the contribution from the region $q
 \lesssim\Omega/v_F \sim T/v_F$
  produces only higher than $T^2$ terms.
In a generic FL, an expansion of the polarization bubble is possible for
  $|\Gamma \Omega_n|/v_F q \ll 1$, where $\Gamma$
  is the largest of the scattering amplitudes in (\ref{32_2}).
   When $\Gamma\gg 1$,
  which
   happens either when the interaction is strong or when the system is near a Pomeranchuk instability,~\cite{mc_10} the
    condition $v_F q \gg |\Gamma\Omega_n| $ sets
    a new lower cutoff for
    integration over $q$.   We consider the case of large $\Gamma$ in Sec.~\ref{sec:4}, where we show that the existence of this cutoff affects the
     prefactor
 for the $T^2$ term in
 Eq.~(\ref{32_2_1}), which  gets smaller as $\Gamma$ increases.

The consequences of the first-Matsubara rule for the de Haas-van Alphen (dHvA) oscillations in a 2D FL were analyzed in Refs.~\onlinecite{martin03,adamov}, where it was shown that the amplitude of these oscillations contains neither a $T^2\ln T$ nor a $T^2$ term
resulting from the self-energy of quasiparticles. This result seems to contradict Eq.~(\ref{32_2_1}) which shows that  the self-energy evaluated at $\omega_m=\pi T$ does have a $T^2$ term. In fact, there is no contradiction because Eq.~(\ref{32_2_1}) refers to the self-energy evaluated on the FS, i.e., for $\ek=0$, while the  dHvA amplitude contains the self-energy evaluated at the \lq\lq Matsubara mass-shell\rq\rq\/, defined by a solution of the equation $G_{\bk}^{-1}(\omega_m)=0$. It turns out that these two self-energies do have different $T$ dependencies.
  The
   amplitude of dHvA oscillations in any thermodynamic quantity contains the following dimensionless combination\cite{wasserman96}
\be
A_{\mathrm{dHvA}}=\frac{iT}{2\pi\omega_c}\sum_{\omega_m>0}\int d\ek G_{\bk}(\omega_m)\exp\left(2\pi i\frac{\ek}{\omega_c}\right),
\label{dhva1}
\ee
where $\omega_c$ is the cyclotron frequency. For simplicity, we omit $\mathcal{O}(\ek)$
and  $\mathcal{O}(\omega_m)$ terms in $\Sigma_\bk$, which only renormalize the effective mass entering the cyclotron frequency, and focus on terms of order $T^2\ln T$ and higher. We also focus on the weak-coupling regime, when the Matsubara mass-shell can be determined perturbatively; to lowest order in the interaction, the mass-shell simply coincides with the pole of the Matsubara Green's function $\ek=i\omega_m$. Substituting the self-energy (\ref{32_2}) evaluated at $\ek=i\omega_m$  into Eq.~(\ref{dhva1}) and integrating over $\ek$, we obtain
\be
A_{\mathrm{dHvA}}=\frac{T}{\omega_c}\sum_{\omega_m>0
}\exp\left(-2\pi \frac{\omega_m+{\tilde \Sigma(\omega_m)}}{\omega_c}\right),
\label{dhva3}
\ee
where
\bea
{\tilde \Sigma}(\omega_m)=-\frac{A_2T}{2\pi }\int^{\Lambda/v_F}_0 dq q\sum_{\Omega_n}\frac{\mathrm{sgn}(\omega_m+\Omega_n)|\Omega_n|}{(v_Fq)^2+\Omega_n^2}.\notag\\
\label{dhva2}
\eea
For high enough temperatures, i..e, for $T\gtrsim\omega_c$, one needs to keep only the $\omega_m=\pi T$ term in the sum of Eq.~(\ref{dhva3}). This is where the first-Matsubara rule becomes useful because the Matsubara sum in Eq.~(\ref{dhva2}) vanishes for $\omega_m=\pi T$, and $A_{\mathrm{dHvA}}$ reduces to the free-electron result (modulo renormalized effective mass) with no extra $T$ dependent terms.

A related point is the difference in the behavior of the self-energy at finite $T$ and at $T=0$.
At $T=0$, the perturbation theory in 2D
for the self-energy
 diverges near the
mass-shell,
 and needs to be
resummed to eliminate these divergences.
\cite{cm_05,cm_03} The mass-shell singularity shows up already in the second-order
self-energy
 at $T=0$, which is obtained by replacing the Matsubara sum in Eq.~(\ref{27_dm}) by an integral over $\Omega_n$. To logarithmic accuracy, this yields
\bwt
\begin{eqnarray}
\Sigma_{\bk} (\omega _{m},T=0) = -\frac{A_2}{8\pi^2v^2_F}\left[ \left( \omega _{m}^{2}+\frac{1%
}{4}~(\omega _{m}+i\varepsilon _{\bk})^{2}\right) \ln \frac{\Lambda}{\omega
_{m}+i\varepsilon _{\bk}}+\left( \omega _{m}^{2}-\frac{1}{4}(\omega_m +i\varepsilon
_{\bk})^{2}\right) \ln \frac{\Lambda}{\omega _{m}-i\varepsilon _{\bk}}\right] .
\label{2.50111}
\end{eqnarray}
\ewt
The mass-shell singularity in this equation is manifested as a divergence of the first logarithmic term at $\ek=i\omega_m$.
However, if we keep $T$  in Eq.~(\ref{27_dm}) finite, integrate over $q$ at finite $\ek$, and then re-arrange the resulting Matsubara sum, we obtain
to logarithmic accuracy and for Matsubara frequencies with $m = O(1)$
\bea
\Sigma_{\bk}(\omega_m,T)&=&-\frac{A_2 T}{2\pi v_F^2}\sum^{\omega_m-\pi T}_{2\pi T}|\Omega_n|\label{ms_2d}\\
&&\times
\ln\frac{\Lambda^2}
{\left(2\Omega_n-\omega_m-i\ek\right)\left(\omega_m+i\ek\right)}.\notag
\eea
The limit of $T\to 0$ in
 this equation reproduces Eq.~(\ref{2.50111}) with the same mass-shell singularity at $\ek=i\omega_m$.
However, at $\omega_m=\pi T$ the sum in Eq.~(\ref{ms_2d}) contains no terms and thus the mass-shell singularity in $\Sigma_{\bk}(\pi T,T)$ is absent. This is the reason why the mass-shell singularity does not show up in the dHvA amplitude.

\subsection{Higher-order terms in canonical Fermi Liquids ($2<D<3$)}

In
canonical
 FLs, the first-Matsubara rule  holds to order $T^2$.
Let us  now verify whether if it also holds to higher orders in $T$.
To obtain $\Sigma_{\bk}(\pi T,T)$ beyond
the
$T^2$
order,
 we need  to go beyond the approximation we
 used in Sec.~\ref{sec:2},
where we assumed that the interaction connected only the points right on the FS.

 We verified that,
 as in 2D,  the terms relevant to our analysis come  both from
  small momentum transfers and momentum transfers near $2k_F$.
Consider for definiteness a small momentum contribution to diagram $a)$ in Fig.~\ref{fig:selfenergy}.
 The corresponding formula for the self-energy is given by  Eq.~(\ref{sigma_a_M}).
 To single out potential terms in $\Sigma_{\bk_F}(
  \pi T,T)$ beyond
 the $T^2$ order, we
  subtract from
   the integrand in Eq.~(\ref{sigma_a_M}) its expression
   for the case when the effective interaction
   connects the points right on the FS.
  We parametrize the measure of the $D$ dimensional integral over $\bq$ as $d^{D-1}q_{\perp} dq_{||}$,
   where a $D-1$ dimensional vector $\bq_{\perp}$ lies in the plane tangential to the FS and $q_{||}$
    is along the normal to the FS, and replace the integral over $q_{||}$ by that over the fermionic dispersion in the final state
    $\epsilon_{
    \bk_F+\bq}\approx v_Fq_{||} \equiv\epsilon$.
  As before, we neglect the static part of $\Pi_{\bq}(\Omega_n)$,
 which contributes only to the
  ${\mathcal O}(T)$ term in $\Sigma$,
    and approximate the dynamic part of $\Pi_{\bq}(\Omega_n)$  by the
    $|\Omega_n|/q=|\Omega_n|/\sqrt{q_\perp^2+q_{||}^2}$ form. Using these simplifications, we express the part of the self-energy not captured
   in Sec.~\ref{sec:2} as
   \bea
&&\delta \Sigma (\pi T,T) =
 - i A_D T
 \sum_{ \Omega_n} | \Omega_n|
 \int \frac{q_{\perp}^{D-2} dq_{\perp} d \epsilon}{(2\pi)^D}
   \nonumber \\
 &&\times  \frac{1}{i(\pi T +  \Omega_n) - \epsilon}
 \left(\frac{1}{\sqrt{v^2_F q_{\perp}^2 + \epsilon^2 + \Omega^2_n}} -\frac{1}{\sqrt{v^2_F q^2_{\perp} + \Omega^2_n}}\right), \nonumber \\
 &&
\label{27_1}
\eea
where
$A_D= 2 \nu_{D} \pi^{\frac{D-1}{2}}/\Gamma[(D-1)/2] U_{\bq=0}^2$,
$\Gamma[x]$ is the Gamma-function,
and $\nu_D$ is the density of states
per spin projection in $D$ dimensions.
Because $\Sigma_{\bk_F} (\pi T,T)$ obtained in Sec.~\ref{sec:2} contains only linear-in-$T$ term and thus satisfies the first-
Matsubara rule, potential deviations from this rule are
 due to $\delta \Sigma (\pi T,T)$.
Integrating over $q_\perp$ and $\epsilon$
 in (\ref{27_1}),
 we find that $\delta \Sigma (\pi T,T)$ contains  a  contribution
\bwt
\bea
\delta \Sigma (\pi T,T) &=&
 A_D \frac{T}{v^D_F} \sum_{\Omega_n} | \Omega_n| (\pi T + \Omega_n) |\pi T + \Omega_n|^{D-3} Q_D \left(\frac{\Omega_n}{\pi T + \Omega_n}\right)\nonumber \\
&& =
 A_D \frac{T^{D}}{v^D_F} \left( 2 \pi\right)^{D-1} \sum_{n=1}^{\bar\Lambda}
 n \left[(n +1/2)^{D-2} Q_D \left(\frac{n}{n+1/2}\right) -(n-1/2)^{D-2}  Q_D \left(\frac{n}{n-1/2}\right) \right],
\label{la_10}
\eea
\ewt
where
${\bar\Lambda}=\Lambda/2\pi T$, and
\bea
&&Q_D (z)\label{u_2}\\
&&=2 \int_0^{\bar \Lambda} \int_0^{\bar \Lambda} \frac{dx x^{D-2} dy}{(2\pi)^D}
 \frac{\sqrt{x^2+y^2 +z^2}-\sqrt{x^2+z^2}}{(y^2+1) \sqrt{x^2+z^2}\sqrt{x^2+y^2+z^2}}.\notag
\eea
The sum in (\ref{la_10})  contains a contribution from the upper limit, which just adds an extra piece to the ${\mathcal O}(T)$ term, but it also contains a $\Lambda$-independent contribution from $n = \mathcal{O}(1)$ which yields $\delta \Sigma (i\pi T, T) \propto T^D$.
We see therefore that the full $\Sigma_{\bk_F} (\pi T, T) = \mathcal{O}(T)+\delta \Sigma (\pi T, T)$ contains
a
$T^D$ term, i.e.,
  the first-Matsubara rule breaks down at order $T^D$ in conventional FLs.  Still, $\Sigma_{\bk_F} (\pi T, T)$ and $\Sigma_{\bk_F} (\omega_m, T)$ for
 $| \omega_m|\neq \pi T$ are qualitatively different:
 the next term after $T$ in $\Sigma_{\bk_F} (\omega_m
 \neq \pi T, T)$ is  $T^2$
 while in $\Sigma_{\bk_F} (\pi T, T)$ it is $T^D$, which for $D >2$ is much
 smaller than $T^2$.
 We
 verified that in the limit $D \to 2$
 the result matches the second term in (\ref{32}). For arbitrary $2<D<3$, the sum has to be evaluated numerically.

The case $D =3$ is
special
 because $Q_3$ diverges logarithmically.
 In this case we have, after integrating over $x$ in (\ref{u_2}) and neglecting non-logarithmic terms,
\bea
&&\delta \Sigma (\pi T,T) \label{u_3}\\
&&= A_3 \frac{4T^3}{\pi v^3_F} \sum_{n} |n| (n+1/2)
  \int_0^{\bar \Lambda} dy \frac{\sqrt{y^2 + z^2}}{y^2 +1},\notag
\eea
where $z = n/(n+1/2)$.
 By power-counting, $\delta \Sigma (\pi T,T)$ scales as $T^3$
 but there is an additional logarithm,
 which can captured by
 expanding
 the summand
 in Eq.~(\ref{u_3})
  in $1/n$.
  The prefactor
  of
    the $T^3$ term is
    $1/|n|$, and the sum $T^3/|n|$
      yields
      a
      $T^3   \ln
   {\Lambda/T}$
   term  in $\Sigma_{\bk_F}(\pi T,T)$. Expanding
  the integrand of Eq.~(\ref{u_3}) in $1/n$,
  integrating
   over $y$, and collecting the prefactors for the $T^3 \ln{\Lambda/T}$
  term,
  we obtain
 \be
\delta \Sigma (\pi T,T) =
 A_3 \frac{T^3}{15 \pi v^3_F} \ln\frac{\Lambda}{T}.
\label{la_10_1}
\ee
The
complete
 result in 3D again contains the contributions from
 diagrams $a)$ and $b)$ in Fig.~\ref{fig:selfenergy} and includes
  terms coming from
 both forward- and backscattering.

Note that the signs of $\delta \Sigma (\pi T, T)$ are different in 2D and 3D [
cf. Eqs.~(\ref{32}) and (\ref{la_10_1})], i.e.,
the prefactor
of the $T^D$ term vanishes at some $D$
in between $2$ and $3$.

\subsection{
Non-canonical Fermi liquids: $1<D<2$}

The analysis for $1<D<2$ parallels that in the previous
 Section.
 The extra term in the self-energy at $\omega_m = \pi T$,
 given by (\ref{la_10}), is
 still
of order $T^D$, and
its
 prefactor is expressed via
 forward-
  and backscattering amplitudes.
 The only difference between
the
 $D<2$ and $D>2$ cases is that, for $D <2$,
 the
 $T^D$ term is larger than
 the
 $T^2$
 one,
 and
 first-Matsubara rule
 breaks down
 completely,
  i.e.,
 the next term after $T$ in $\Sigma_{\bk_F}(\omega_m, T)$ is of order $T^D$ for all
 $\omega_m$ including $\omega_m=\pm\pi T$.

\section{The first-Matsubara-frequency rule near quantum criticality}
\label{sec:4}
 \subsection{Local approximation}
 \label{sec:4a}

So far we found that in a generic FL, either conventional or unconventional, the terms of order $T^D$ in the self-energy
 do not
 distinguish
  between
  the
   first and other Matsubara frequencies,
 i.e., the prefactor
of
  the $T^D$ term in $\Sigma_{\bk_F} (\omega_m)$
  is non-zero for all $m$.

We now show that a different situation emerges when
the system
is tuned to
the
vicinity of a Pomeranchuk transition, at
 which a FL becomes unstable towards condensation of particle-hole excitations with zero momentum transfer.  A Pomeranchuk instability
 can  occur in either
 the spin or
 charge channel.
   A magnetic (spin)  instability
   is
   likely
   to trigger
   pre-emptive transitions,~{\cite{belitz,mc_inst} and, to keep
   the discussion
    focused on the first-Matsubara rule, we only consider here
  a Pomeranchuk instability in the charge channel.   In the bulk of this section we focus on long-wavelength ($q=0$) Pomeranchuk instability,
  (a quantum phase  transition with dynamical exponent $z=3$).  At the end of this section, we briefly discuss the first-Matsubara rule near an
   instability  at
   finite $q$ in a system on
   lattice (a quantum phase  transition with dynamical exponent $z=2$).

Near a Pomeranchuk  instability,
 interactions generate a large length scale $\xi$ (the correlation length) which diverges at
 the transition. In
 $D \leq 3$, a
 divergence in $\xi$ brings the upper
 boundary
 of FL behavior down from $\mathcal{O}(E_F)$ to  $\omega_{\mathrm{FL}}
 \propto \xi^{-3}$.\cite{mc_10}
  At large enough $\xi$, $\omega_{\mathrm{FL}}$  becomes smaller than
  $\Lambda$,
  and the low-energy theory with the upper cutoff  $\Lambda$
  describes now both the FL and non-FL regimes. We first consider
   the case of
  $\Omega, T
  \ll\omega_{\mathrm{FL}}$ and then
   discuss the first-Matsubara rule at energies above $\omega_{\mathrm{FL}}$.

 The observation,
 which is
  most relevant to our analysis, concerns the low-energy cutoff in the integration over bosonic momentum $q$ in
  the  formula for the self-energy, once we cast it into the form of
  Eq. (\ref{27}).
  As we
  mentioned  in Sec.~\ref{sec:d=2},
  the $T^D$ term  in $\Sigma_{\bk_F}(\omega_m)$
  with a prefactor that does not show any special features at $m=0,-1$
  is obtained by
  setting
   the lower momentum cutoff to zero.  This approximation can be justified at $\xi =
    \mathcal{O}(1)
    \sim k_F^{-1}
    $, at least at weak coupling,
    but not at large $\xi$.  To  show this, we follow earlier work~\cite{nematic}
     and assume that, near a Pomeranchuk instability
    with some angular momentum $
    \ell$,  the fermionic self-energy
     given Eq. (\ref{27})
     can be
     viewed as
     resulting
     from
     an exchange of low-energy
     and overdamped collective excitations.
     The propagator of these excitations at small $q$ is given by
     \be
     \chi_{\bf q}(\Omega_n) = \frac{\chi_0}{q^2 + \xi^{-2} + \gamma \Pi_{\bf q} (\Omega_n)},
     \label{m_1}
     \ee
    where  $\gamma$
    depends on
    original fermion-fermion interaction and fermionic dispersion
    and,
    in general,
     is different for different $
     \ell$.
      As
       before, we  keep only the dynamic part in $\Pi_{\bq} (\Omega_n)$.

         The
         one-loop
     self-energy
      is given by
\be
\Sigma_{\bk_F} (\omega_m,T) =
i
T \sum_{ \Omega_n}  \int \frac{d^D q}{(2\pi)^{D}} G_{{\bf k}_F +
 {\bq}} (\omega_m + \Omega_n) \chi_{\bq} (\Omega_n).
\label{m_2}
\ee
 An order-of-magnitude estimate for $\Sigma_{\bk_F} (\omega_m,T)$ can be obtained by expanding  $\chi_{\bq}(\Omega_n)$ as
\be
\chi_{\bq} (\Omega_n) = \chi_{\bq} (0) -
\chi_0
\gamma \xi^4 \Pi_{\bq} (\Omega_n).
\label{m_3}
\ee
Substituting this expansion into
Eq.~(\ref{m_2})
 and comparing the result to (\ref{27}) in Sec. ~\ref{sec:d=2},
and to its extension for
 an
 arbitrary interaction in Eq.~(\ref{32_2}),  we
 see that
 $\chi_0\gamma \xi^4$ plays the same role as
 the combination of the
 $\Gamma^2$ terms in
Eq.~(\ref{32_2}), i.e., the overall  prefactor
 of the dynamic part
of
 $\Sigma_{\bk_F} (\omega_m,T)$ scales as $\xi^4$.

Let us now look more carefully
 at
the
 limits of integration over $q$,
 which
need to be imposed to ensure self-consistency expansion (\ref{m_3}) for $\chi_{\bq}$.
  Because $\Pi_q (\Omega_n)$ scales as $|\Omega_n|/q$ at the smallest $\Omega_n$,
 the expansion in $\Pi_{\bf q} (\Omega_n)$ in (\ref{m_3}) holds only for $q > \gamma |\Omega_n| \xi^2$,
 which sets the lower
 limit
  in the
  integral
   over $q$. The upper
   limit
    is set  by $\xi^{-1}$.
   Now,  we expand the
       dispersion as
     $\varepsilon_{\bk_F+\bq}=v_Fq_{||}+q^2_{\perp}/2m^*
     $ and express $q_{||}$
        as
     $q_{||}= \varepsilon_{\bk_F+\bq}/v_F-q_{\perp}^2/(2 v_F
     m^*)
    $.
 Consider
momentarily a free fermion propagator
in (\ref{m_2}). Typical $\varepsilon_{\bk_F+\bq}$ are then of order $\omega_m + \Omega_n$, i.e.,
 of order $T$ for Matsubara indices $m,n\sim 1$.
 Since we expect the $T^D$ term to come from the region when both typical $q_{||}$ and $q_{\perp}$ are also
proportional to $T$,
 the $q_\perp^2/2m^*$ term is of order $T^2$ and can be neglected compared to $\varepsilon_{\bk_F+\bq}$.
 Hence typical
 $q_{||}
 \sim
  \varepsilon_{\bk_F+\bq}/v_F$, and typical
 $q
 \sim
  \sqrt{q^2_\perp + \left[(\omega_m + \Omega_n)/v_F\right]^2}$.
  For
   large $\xi$ and $
   n,m
   \sim 1$,
    $(\omega_m + \Omega_n)/v_F \sim \Omega_n/v_F$ is then smaller by
  than the lower
  limit
  for $q$, which is  $\gamma |\Omega_n| \xi^2$. For $\gamma v_F  \xi^2 \gg 1$, one can
  approximate $q$ by $q_\perp$.
     This
      is equivalent to factorizing the momentum integral in
  (\ref{m_2})
  as
   $\int d \epsilon_{{\bf k}_F+{\bf q}} G_{{\bf k}_F+{\bf q}} (\omega_m + \Omega_n) \int d^{D-1} q_{\perp} \chi_{\bq_\perp} (\Omega_n)$. In this approximation, the
   dynamic part of the
   self-energy in (\ref{m_2}) reduces to
  \be
  \Sigma _{\bk_F}(\omega_m,T) =
  \frac{T}{2v_F}
  \sum_{ \Omega_n}  \mathrm {sgn}(\omega_m + \Omega_n) \chi_L (\Omega_n),
\label{m_2_1}
\ee
 where
  \be
 \chi_L (\Omega_n)= \int \frac{d^{D-1} q_\perp}{(2\pi)^{D-1}}~ \chi_{\bq_\perp} (\Omega_n).
 \ee
 Because $\chi_L (\Omega_n)$ is an even function of $\Omega_n$, the r.h.s. of (\ref{m_2_1}) vanishes at
$|\omega_m|=\pi T$,
 i.e.,  the first-Matsubara rule holds.
  For all other frequencies, such that
 $|\omega_m|\neq \pi T$ but still
 $|\omega_m|\sim T$,
   $\Sigma_{\bk_F}(\omega_m,T)$  behaves as $T^{D} \xi^{2D}$ in non-conventional FLs, and   as
  \bea
  \Sigma_{\bk_F} (\omega_m,T)
  =
   T^2 \xi^{6-D}
  \left\{1 +\mathcal{O}\left(\left(T\xi^3\right)^{D-2}\right)\right\}
  \label{h_1}
  \eea
  in conventional FLs.

  We
  refer to an
   approximation, in which the momentum integral is factorized, as
   the \lq\lq local approximation\rq\rq\/. The name reflects the fact that
   the fermionic self-energy in this approximation
   is a convolution
   of the
   density of states (the Green's function integrated over fermionic dispersion) and
   the {\em local}
 susceptibility,
obtained by integrating the non-local susceptibility over $D-1$ components of $\bq_{\perp}$.

  If we keep $\omega_m + \Omega_n$ in $q$ and compute $\Sigma_{\bk_F}(\omega_m,T)$ without making any approximations, we find that the
  $T^D$ term
  in $\Sigma_{\bk_F}(\omega_m,T)$  is present  for all $
  \omega_m
  $; however,
  its
  prefactor has different dependences on $\xi$ for
  $|\omega_m|=\pi T$ and all other $\omega_m$.
  For
  $|\omega_m|\neq \pi T$,
  the  $T^D$
    term in $\Sigma_{\bk_F} (\omega_m,T)$ is present even in the local approximation,
    and the prefactor of
    this term scales as $\xi^{2D}$. For
    $|\omega_m|=\pi T$,
    the prefactor is zero in the local approximation, and scales as $\xi^{2(D-2)}$ if we compute $\Sigma_{\bk_F} (\pi T,T)$ in (\ref{m_2}) using a free fermion propagator. Using free-fermion propagator at large $\xi$ is, however, not
    justified
    because the mass renormalization  term,
   $\lambda \omega_m$,
     in $\Sigma_{\bk_F}$
     is also proportional to $\xi$
     [this term involves a static susceptibility,
     $\chi_L (0)$].  Including this term into the Green's function
    affects the estimate for a typical $\varepsilon_{{\bf k}_F + {\bf q}}$, which now becomes of order $(1 + \lambda) |\omega_m + \Omega_n|/v_F$.
    Accordingly, the prefactor of the $T^D$ term  scales as $\lambda^2 \xi^{2(D-2)}$ at
    $|\omega_m|=\pi T$.
    At one-loop order, $\lambda \propto \xi^{3-D}$ ($\propto \ln \xi$ in $D=3$) and, hence, the self-energy at the first Matsubara frequency scales as
    $\Sigma_{\bk_F} (\pi T,T) \propto T^D \xi^{2}$. Still, for all $D >1$, this is parametrically smaller than the self-energy at larger Matsubara frequencies, which, we remind,  scales as $T^2 \xi^{6-D}$ in conventional FLs, and as $T^D \xi^{2D}$ in non-conventional FLs.

    The
    main
    outcome of this analysis is that,
     near a Pomeranchuk instability
          the first-Matsubara rule  approximately holds, even if far from the instability
     this rule
     is
     broken, as it happens in  non-conventional FLs.
 The distinction between the prefactors of $T^D$ terms in $\Sigma _{\bk_F}(\omega_m,T)$ likely persists
 to higher-orders in the
 loop expansion,
  even if higher-order corrections are not small. To verify this, we analyzed two and three-loop contributions to the self-energy near a charge Pomeranchuk transition in $D=2$. We recall that in $D=2$, the self-energy at a generic $\omega_m$ scales as $T^2 \xi^4$.
  In 2D,  a two-loop self-energy is small compared to
 the  one-loop
  one,
   Eq.~(\ref{m_2}), only if one extends the theory to $N$ fermionic flavors and takes the $N \gg 1$ limit.~\cite{aim,rech}
  The
   three-loop self-energy is not  small even
  in the
  large-$N$
  limit
  (Refs.~\onlinecite{sslee, metl_sachdev,senthil}), and higher-order terms even bring in additional logarithmic singularities.~\cite{metl_sachdev,senthil,ckl}  We computed two-loop and three-loop contributions to the self-energy
   along Matsubara axis, and found that in both contributions the prefactor for the $T^2$ term still vanishes at $\omega_m = \pm\pi T$ if the local approximation is imposed, and scales as $\xi^2$
   beyond this approximation.
    Higher-order corrections may, in principle, generate additional logarithms
    and eventually change the
    scaling
     of $\Sigma_{\bk_F}(\pi,T)$ with $\xi$ from $\xi^2$ to  $\xi^\beta$ with $\beta<2$.
    However, because
    the one-loop results for
    $\Sigma_{\bk_F} (\pi T,T)$ and $\Sigma_{\bk_F}(\omega_m,T)$ with $|\omega_m| \neq \pi T$
    differ substantially (by
  a factor of  $\xi^2$ in $D=2$),
  it is
  likely
   that
   the
    difference between the prefactors of $T^D$ terms in $\Sigma_{\bk_F}(\pi T,T)$ and in
   $\Sigma_{\bk_F}(|\omega_m|\neq \pi T,T)$
   holds  to infinite order in
   the loop-expansion.

The difference
between $\Sigma_{\bk_F} (\pi T,T)$ and $\Sigma_{\bk_F}(
|\omega_m|\neq \pi T,T
)$
 becomes particularly
 pronounced
 right
 at the
 Pomeranchuk instability.  Now $\omega_{FL} =0$,
 and the self-energy exhibits a non-FL  behavior at any finite $\omega$ or $T$.
 The self-energy for generic $\omega_m\neq\pm \pi T$ can be divided into two parts:
 dynamic, $\Sigma^{\mathrm{d}}$, and static, $\Sigma^{\mathrm{s}}$. The dynamic part comes from processes with non-zero energy transfers,
 corresponding to $\Omega_n\neq 0$ in the Matsubara sum of Eq.~(\ref{m_2}).
 The critical form of the dynamic part is obtained by replacing $\xi^{-3}$ in Eq.~(\ref{h_1}) by
 $T$, which gives
  $\Sigma^{\mathrm{d}}_{\bk_F} (\omega_m,T) \sim  T^{D/3}$.
  The static part comes from scattering of static critical fluctuations, corresponding to a single term with $\Omega_n=0$ in
  Eq.~(\ref{m_2}).
    At
     finite $\xi$, this contribution
    behaves as $\Sigma^{\mathrm{s}}_{\bk_F}(T)\propto T\xi^{3-D}$. At $\xi\to\infty$, the static contribution diverges for $D\leq 3$.\cite{acs,millis94}
  This divergence is usually regularized by introducing a temperature-dependent correlation length, $\xi(T)$, which remains finite at   $T >0$  even right at criticality. On general grounds, one can postulate that $\xi(T)\propto T^{-\beta_T}$ with $\beta_T>0$,
   and hence  $\Sigma^{\mathrm{s}}_{\bk_F}(T)\propto T^{1-\beta_T(3-D)}$.
   At
   the
   one-loop level, $\beta_T=1/2$
   (modulo logarithms) for 2D
   quantum-critical
   systems with
    dynamical exponents
   $Z=2$ and $Z=3$, \cite{millis94}
   but higher-order corrections may change the exponent. We
 will treat $\beta_T$ as a phenomenological parameter of the theory. Comparing the exponents of the dynamic and static parts of the self-energy, we see that, for any $D<3$,  the leading $T$ dependence of the self-energy is given by the dynamic part  if
 $\beta_T<1/3$
  and by the static part if
  $\beta_T>1/3$.

  For the first Matsubara frequency, the static part of the self-energy is the same as for all other $\omega_m$, but the dynamic part is different.
  To obtain
  $\Sigma^{\mathrm{d}}_{\bk_F} (\pi T,T)$ at criticality, we re-evaluate
  the self-energy diagram in (\ref{m_2}) by replacing  the
frequency in the denominator of the Green's function
by the self-energy at the same frequency. Now
 typical
$\varepsilon_{\bk_F+\bq}\sim \Sigma_{\bk_F}(\omega_m+\Omega_n,T)\left\vert_{\omega_m+\Omega_n\sim T}\right.\equiv  {\bar\Sigma} (T)$.
Expanding the bosonic propagator to leading (second) order in $\varepsilon_{\bk_F+\bq}$ and performing power-counting, we obtain
\bea
\Sigma^{\mathrm{d}}_{\bk_F}(\pi T,T)\propto T^{\frac{D-2}{3}}{\bar\Sigma}^2\propto
\left\{
\begin{array}{cc}
T^{D-2/3}\;
\mathrm{, if}\;\beta_T
<1/3\\
T^{\frac{D+4}{3}-2\beta_T(3-D)}\mathrm{, if}\;\beta_T
>1/3
\end{array}
\right.
,\notag\\
\eea
where we replaced ${\bar\Sigma}$ by $\Sigma^{\mathrm{d}}$ and $\Sigma^{\mathrm{s}}$ for $\beta_T
<
1/3$ and $\beta_T>1/3$, correspondingly.
 We see that $\Sigma^{\mathrm{d}}_{\bk_F}(\pi T,T)$ remains smaller than $\Sigma^{\mathrm{d}}_{\bk_F}(\omega_m\neq\pi T,T)$: the ratio of the two
 behaves as
$\Sigma_{\bk_F}(\pi T,T)/\Sigma^{\mathrm{d}}\propto T^{2(D-1)/3}$
for $\beta_T
<1/3$,
 and $\Sigma_{\bk_F}(\pi T,T)/\Sigma^{\mathrm{d}}\propto T^{(D+1)/3-\beta_T(3-D)}$ for $\beta_T
 >1/3$.
 The exponent is positive for any $D>1$  in the first expression  and for $1/3<\beta_T<(D+1)/3(3-D)$ in the second one.
    In both cases the ratio of the self-energy at the first Matsubara frequency to that at a generic frequency scales to zero as $T$ goes to zero. This smallness is a manifestation
of the first-Matsubara rule at criticality.

\subsection{Scaling form of the self-energy in the local approximation}

 A non-trivial aspect of
 the
 first-Matsubara rule near a Pomeranchuk instability shows up when we consider the self-energy
  along
  the
  real frequency axis. At
  order $T^D$,
   both
   the
  real and imaginary parts of $\Sigma^R_{\bk_F} (\omega, T)$  are rather
  complicated
   functions of $\omega$
   and
   $T$, and the
 extension of $\Sigma^R
 _{\bk_F} (\omega, T=0)$ to
  finite $T$
  by no means implies that $\omega$ is replaced by
  $\sqrt{\omega^2 + \pi^2 T^2}$. Still,
  within
  the
  local approximation, we obtain,
 analytically continuing
  (\ref{m_2_1}) to real frequencies
\be
\mathrm{Im} \Sigma^R_{\bk_F} (\omega,T) =
 \frac{1}{2\pi v_F}
\int d
\Omega \mathrm{Im} \chi_L ^R(
\Omega) \left[n_B (
\Omega
) + n_F (
\Omega+\omega)\right].
\label{38}
\ee
The r.h.s. of Eq.~(\ref{38})  is an analytic
function of complex variable
$\omega\to z=z'+iz''$ within the stripe $
|\I z|
\leq \pi T$, see Fig.~\ref{fig:bcuts}.
 Within this stripe,  one can then
 analytically continue $\mathrm{Im} \Sigma_{\bk_F}^R (\omega,T)$ into the complex plane by just replacing $\omega \to z$.
  Aa a result, $\mathrm{Im} \Sigma^R_{\bk_F}( i\pi T, T)$
 is still given by (\ref{38}), but with $i\pi T$ instead of $\omega$ in the r.h.s of this equation.  Because $n_B (\Omega) + n_F (\Omega+i\pi T) =0$,
$ \mathrm{Im}\Sigma^R
_{\bk_F}
 (\omega =i\pi T, T)$  vanishes.  The full $\Sigma_{\bk_F}
 ^R
 (i\pi T, T)$ vanishes by
 the
  first-Matsubara rule, hence $\mathrm{Re}
   \Sigma
   ^R_{\bk_F}
   (\omega, T)$
 must also vanish (up to
 a $\mathrm{O}(T)$ term),
  if we replace $\omega$ by $i \pi T$.  These two requirements then set  non-trivial constraints on
  the scaling functions of $\omega/T$ in  $ \mathrm{Im} \Sigma^R_{\bk_F} (\omega,T) \propto |\omega|^
  {D} f_{ID} (|\omega|/T)$ and
  $ \mathrm{Re} \Sigma^R_{\bk_F} (\omega,T) \propto \omega |\omega|^{D-1}
  f_{RD} (|\omega|/T)$:  both $f_{ID} (x)$ and $f_{RD} (x)$ must vanish
  at
  $x = i \pi$.

In the next two
 sections, we obtain explicit forms of $f_{ID} (|\omega|/T)$ and $f_{RD} (|\omega|/T)$ for
near-critical FLs in $D =2$ and $D=3$ and show they they satisfy the constraint.\\

\subsubsection{$D=2$}

We again use (\ref{m_1}) for $\chi_q (\Omega_n)$.
  In $D=2$ we have
 \be
 \I \chi_L^R (\Omega) = \frac{
    \chi_0
   \gamma \xi^4}{\pi v_F} \Omega \ln{\frac{\omega_{\mathrm{FL}}}{|\Omega|}},
 \label{40}
 \ee
  where $\omega_{\mathrm{FL}}
  \sim 1/
  (\gamma \xi^3) $ is the upper boundary of the FL behavior.
  The imaginary part of the self-energy is given by
 \be
\mathrm{Im} \Sigma^R
_{\bk_F}
 (\omega,T) =
 B_0
 \int d \Omega \Omega \ln{\frac{\omega_{FL}}{|\Omega|}} \left(n_B (\Omega) + n_F (\Omega+\omega)\right),
\label{41}
\ee
 where
  $B_0 = \chi_0\gamma \xi^4/(2\pi^2 v^2_F)$.
 The real part of the self-energy is obtained
 via the
 KK  relation.
 We skip the details of calculations and show only the final results. It turns out that the real part of the self-energy (the one which does not contain logarithms) can be computed exactly,
 up to the term of order $\omega$ which we omit below.
   The real part of the self-energy is
   an
   odd function of
   the
   frequency at
   $\bk=\bk_F$. For
  $\omega >0$ we find
 \bea
 &&\mathrm{Re} \Sigma^R_{\bk_F} (\omega,T)
 \label{42}
\\
  &&
  =
  -\frac{B_0}{4} \left[\pi \omega^2 + 4
  \pi T^2
  \left(\frac{\pi^2}{12} + \mathrm{Li}_2 \left[-e^{-
  \omega/T}\right]\right)\right],\nonumber
\eea
where
\be
\mathrm{Li}_s (y) =
\sum_{k=1}^{\infty} \frac{z^k}{k^s}\ee
 is
 a
 polylogarithmic function.
 This expression can be cast into the scaling form  $\mathrm{Re} \Sigma^R_{\bk_F} (\omega,T) = \omega|\omega| f_{R2} (|\omega|/T)$.
 For $\omega = i\pi T$,  $-e^{-\omega/T} =1$ and $\mathrm{Li}_2(1) = \pi^2/6$.  Substituting these relations into (\ref{42}) we find that $\mathrm{Re}\Sigma^R_{\bk_F}( i\pi T,T) =0$, as expected.

The imaginary part of the self-energy is given by Eq. (\ref{41}) in the form of a one-dimensional integral.
The formula for $\mathrm{Im} \Sigma^R_{\bk_F} (\omega,T)$ can be simplified if we extract from it the leading logarithmic term.
Combining the remainder of $\mathrm{Im} \Sigma^R_{\bk_F} (\omega,T)$ with $ \mathrm{Re} \Sigma_{\bk_F}^R(\omega,T)$,
we obtain
\begin{widetext}
\bea
  \Sigma^R_{\bk_F} (\omega,T) &=& i \frac{
  B_0}{4} \left(\omega^2 + \pi^2 T^2\right) \ln{\frac{e (\omega_{\mathrm{FL}})^2}{\pi^2 T^2}}
 + i\frac{
 B_0
 }{4} \left(\omega^2 + \frac{\pi^2 T^2}{3}\right) \ln{\frac{\pi^2 T^2}{-\omega^2}} \nonumber \\
 &&+ 2i
 B_0
  T^2
 \int_0^\infty x \mathrm{Li}_2(-e^{-\pi x}) \left(\frac{1}{x^2 -(\omega/\pi T)^2} -\frac{1}{x^2 +1} \right),
\label{43}
\eea
\end{widetext}
where
 $\ln (-\omega^2) = \ln{\omega^2} -i\pi
 \text{sgn} \omega $.
In
Eq.~
(\ref{43}),
we singled out
 the leading, logarithmic term in $\mathrm{Im} \Sigma^R_{\bk_F} (\omega,T)$
 and the rest has the form $\omega^2 f_{I2}(|\omega|/T)$. The scaling function is  rather non-trivial, yet
 we see from (\ref{43}) that $\Sigma^R_{\bk_F} (i\pi T,T)$ vanishes, as it should.

 In Appendix \ref{app_subtle}, we discuss several subtle issues related to
  analytic continuation of the self-energy to complex $\omega$ plane in a situation when either $ \mathrm{Re} \Sigma^R_{\bk_F} (\omega,T)$  or $ \mathrm{Im}\Sigma^R_{\bk_F} (\omega,T)$ cannot be evaluated explicitly and
 has to be kept in an integral form, as in (\ref{41}).

\subsubsection{Subleading terms in $D=3$}

A very similar situation emerges in 3D systems
 if we go beyond the leading, $\omega^2 + \pi^2 T^2$ term in the self-energy and consider the subleading terms of order $T^3$ and $\omega^3$.
 At $T=0$,
   the real part of the self-energy scales as $\omega^3 \ln|\omega|$
 and the imaginary part scales as $|\omega|^3$. At finite $T$,
 both
 parts
 contain scaling functions of $|\omega|/T$.
 The situation is somewhat similar to that in $D=2$ in
 a sense that the behavior is marginal due to logarithms.

 Using
 $\chi_{\bq} (\Omega_n)$
 from Eq.~(\ref {m_1})
   we obtain
 \be
\I\chi_L^R
 (\Omega)=  \frac{\chi_0 \gamma \xi^3}{8v_F} \Omega -
\frac{\chi_0 \gamma^2 \xi^6}{4 \pi v^2_F} \Omega |\Omega| + \cdots
\label{45}
\ee
Substituting this form into Eq.~(\ref{38}),
 we obtain after some algebra
an
 explicit
 expression
  for $\mathrm{Im} \Sigma^R_{\bk_F} (\omega,T)$
 to order $\omega^3, T^3$:
\bwt
\bea
&&\mathrm{Im} \Sigma^R_{\bk_F} (\omega,T)
\nonumber \\
 &&= C_0\left(\omega^2 +\pi^2 T^2\right) +
D_0 \left\{\frac{|\omega|}{3} \left(\omega^2 +\pi^2 T^2\right) + 4 T^3 \left(\mathrm{Li}_3 (-e^{|\omega|/T})- \zeta(3)\right)\right\},
\label{46}
\eea
\ewt
where
$C_0= \chi_0 \gamma \xi^3/(32\pi v^2_F)$, $D_0 = \chi_0 \gamma \xi^6/(24\pi^2 v^3_F)$, and
 $\zeta (
 x)$ is
 the
 zeta function.
 Using
 that
 $\mathrm{Li}_3 (1) = \zeta (3)$,
 one can immediately verify that  $\mathrm{Im} \Sigma^R_{\bk_F} (i\pi T, T) =0$, as it should.  This happens despite that
 the functional form  $ \mathrm{Im} \Sigma^R
 _{\bk_F}
  (\omega,T)$ is rather
 complicated
  at order $\omega^3, T^3$;
    e.g., the prefactor
    of the $\omega^3$ term
    is not the same as the prefactor
    of the $T^3$ term.

 The real part of the self-energy
 contains logarithms
  and has to be left in
  an integral form.
  The calculation of
$ \mathrm{Re}\Sigma^R_{\bk_F} (\omega,T)$ using
the
 KK formula, Eq.~(\ref{11}),  requires some care as the integral is formally infrared divergent, if we use Eq.~(\ref{46}) for $\mathrm{Im}\Sigma^R_{\bk_F} (\omega,T)$. The recipe is to i) start with the general expression for $\mathrm{Im}\Sigma^R_{\bk_F}(\omega,T)$ in  Eq.~(\ref{38}); ii) substitute it into
the KK formula and obtain $ \mathrm{Re} \Sigma_{\bk_F}^R
 (\omega,T)$ in the form of
 a double integral;
 iii) keep the full
 form of
 $\mathrm{Im}\chi_L^R
 (\Omega)$ (without expanding it) at intermediate stages of
 the
  calculation,
  and change the order of integrations when it is convenient, iv) use the fact that
 $ \mathrm{Im}\chi_L (\Omega)$ vanishes in the infra-red and also
  that
  $
  \R  \chi_L^R (0) = (2/\pi) \int_0^\infty d\Omega
  \mathrm{Im}\chi_L
  ^R
  (\Omega)/\Omega$. Evaluating  $\mathrm{Re}\Sigma^R_{\bk_F}(\omega,T)$ this way, we obtain
    \bwt
  \bea
  && \mathrm{Re}\Sigma^R_{\bk_F}(\omega,T) = \omega \frac{
  \R \chi^R_L (0)}{2\pi v_F}
 + \frac{1}{3\pi^2 v_F} \omega (\omega^2 + \pi^2 T^2) \int_0^\infty
  \mathrm{Im}\chi_L
  ^R
  (x) \frac{x dx}{(x^2 - \omega^2)^2}
  \nonumber \\
  &&+
   \frac{\omega}{\pi^2 v_F} \int_0^\infty
  \mathrm{Im}\chi_L
  ^R
  (x) dx \left[\frac{1}{2\omega} \ln{\frac{x+\omega}{x-\omega}} - \frac{1}{x} - \frac{x \omega^2}{3(x^2-\omega^2)^2}\right]
  \nonumber \\
  && +\frac{4\omega}{\pi^2v_F} \int_0^\infty
  \mathrm{Im}\chi_L^R(x) x dx \int_0^\infty \frac{y dy}{e^{x/T} +1} \left[\frac{1}{(x^2+y^2 -\omega^2)^2 -4 x^2y^2} -
  \frac{1}{(x^2-\omega^2)^2}\right],
  \label{47}
  \eea
  \ewt
  where  all integrals are
  to be
   understood
 as principal values.
 The last two integrals are ultra-violet convergent
  for $ \mathrm{Im}\chi_L^R(x)$
  given by (\ref{45}). The second term
   in (\ref{47})
   is  singular, but only logarithmically, and accounts for
 the
  $\omega^3 \ln |\omega|$ term in
$ \mathrm{Re}\Sigma^R_{\bk_F}$.
 Substituting  $\chi_L ^R(\omega)$ from
 Eq.~(\ref{45}) and combining $\mathrm{Im}\Sigma^R_{\bk_F}(\omega,T)$ and $\mathrm{Re}\Sigma_{\bk_F} (\omega,T)$,
 we obtain from (\ref{47})
\bwt
 \bea
&& \Sigma^R_{\bk_F}(\omega,T) = \omega \frac{
\R
\chi
^R_L (0)}{2\pi v_F}
 +  i C_0 \left(\omega^2 + \pi^2 T^2\right) +
\frac{D_0}{\pi} \omega \left(\omega^2 + \pi^2 T^2\right) \ln\left[{\frac{\omega^2_{\mathrm{FL}}}{-e \omega^2}}\right] + \frac{5 D_0\omega^3}{3\pi}
\nonumber \\
&& -\frac{4D_0}{\pi} \omega \int_0^\infty \frac{dx}{x^2 + \omega^2} \left[x(x^2 + \pi^2 T^2) + 6 T^3\left(\mathrm{Li}_3 (-e^{x/T}) - \mathrm{Li}_3 (1)\right)\right].
\label{48}
\eea
\ewt
 Equation (\ref{48}) is
 a complete
 expression for the self-energy in a 3D FL within the local approximation.

 One can easily make sure that $\Sigma^R_{\bk_F}(\omega, T)$ in (\ref{48}) is an analytic function of $\omega$ in the
  upper half-plane,
   hence
 it can be straightforwardly
 continued
  from the real axis
  into
  the upper half-plane just by replacing $\omega$ by a complex $z$. At $z = i\pi T$, the second and third term vanish,
  while
   the last two terms cancel each other, i.e., at the first Matsubara frequency the self-energy contains a linear in $T$ term but no terms of higher power of $T$,
   in agreement with the first-Matsubara rule.

\subsection{Marginal FL}

As another illustration,
we consider
the
self-energy in a marginal FL (MFL).\cite{mfl}  The term marginal FL refers to a situation when the imaginary part of the self-energy is comparable to $\omega$, hence by
 the Landau criterion,
 the
 system is at the boundary between FLs and non-FLs.
 By
  the
 KK relation, if
  $\I \Sigma (\omega, T=0)
 \propto|\omega|$, then $\R\Sigma (\omega, T=0)
  \propto \omega \ln( \Lambda/|\omega|)$.

 Because in a
  generic non-conventional FL  $Im\Sigma^R(\omega,T) \propto \omega^{D} f_{ID} (|\omega|/T)$,
 the
  MFL behavior formally emerges when $D$ approaches one. This limit is, however, special, and below we follow earlier work
 \cite{abrahams:2000}
  and assume that
 the
 MFL behavior is associated with some sort of quantum criticality rather than with $D=1$.
  Specifically, the MFL behavior emerges
  if one
 assumes
  $\I\chi_L^R(\Omega, T)$ to be a scaling function of $\Omega/T$
  such that $\I\chi_L^R(\Omega, T=0)=\mathrm{const}\times\mathrm{sgn}(\Omega)$
  and $\I\chi_L^R(\Omega, T)\propto \Omega/T$ for $\Omega\ll T$.
  \cite{abrahams:2000}
  A simple model form of  $\I\chi_L^R (\Omega, T)$ satisfying these conditions is
 \be
 \I\chi_L^R (\Omega, T) =
 \chi_{L0} \tanh {\frac{\Omega}{T}}
 \label{h_4}
 \ee
  This expression is valid for $\Omega$ smaller than some
  cutoff
  energy
    $E^*$.
    At larger $\Omega$, $\I\chi_L^R(\Omega, T)$ must
    decrease.
     To simplify calculations, we impose a hard cutoff, i.e., set $\I \chi_L^R (\Omega, T)$ to be given by (\ref{h_4}) for $|\Omega| <E^*$ and
    $\I \chi_L ^R(\Omega, T) =0$ for $|\Omega| >E^*$.

   The first-Matsubara rule states that the self-energy at the first Matsubara frequency must be
   $\Sigma_{\bk_F} (\pi T,T) =
    \pi T
    \chi_L (0,T)/
    (2\pi v_F)
    $.
   In
   all
   examples
   considered
   so far,
   we assumed that $D>1$
   and hence
   dropped this term, as
    it was of different order than the $T^D$ term which was  our primary interest.
   Now $\I \Sigma^R_{\bk_F}(\omega \sim T, T) = \mathcal{O}(T)$, and we should keep all $\mathcal{O}(T)$ terms.

 Substituting Eq.~(\ref{h_4}) into
 Eq.~(\ref{38}),
 we obtain
 \be
\mathrm{Im} \Sigma^R_{\bk_F} (\omega,T) = \frac{\chi_{L0}}{2\pi v_F} \int d \Omega \tanh\frac{\Omega}{T} \left[n_B (\Omega) + n_F (\Omega+\omega)\right].
\label{h_2}
\ee
Because the integral converges at large $\Omega$, and we are interested in $\omega, T \ll E^*$, we can safely extend  integration over $\Omega$ to the whole real axis. At $T=0$ we have from (\ref{h_2}) $\mathrm{Im}\Sigma^R_{\bk_F}(\omega,0) = \chi_{L0}\omega/2\pi v_F$, and at $\omega=0$,  $\mathrm{Im} \Sigma^R_{\bk_F} (0,T) = \chi_{L0}\pi T/2\pi v_F$.
 When $\omega$ and $T$ are both  finite,
 integration in (\ref{h_2}) yields
 \bea
 &&\mathrm{Im} \Sigma^R_{\bk_F} (\omega,T) = \frac{T\chi_{L0}}{2\pi v_F} f_{IM} \left(\frac{\omega}{T}\right) \nonumber \\
 && f_{IM} (x) = \frac{(\pi/2)(e^x+1)^2 + x (e^{2x}-1)}{e^{2x} +1}.
\label{h_3}
\eea
Function $f_{IM}(x)$ is plotted in Fig.~\ref{fig:mfl}.
 \begin{figure}[t]
\includegraphics[width=0.5\textwidth]{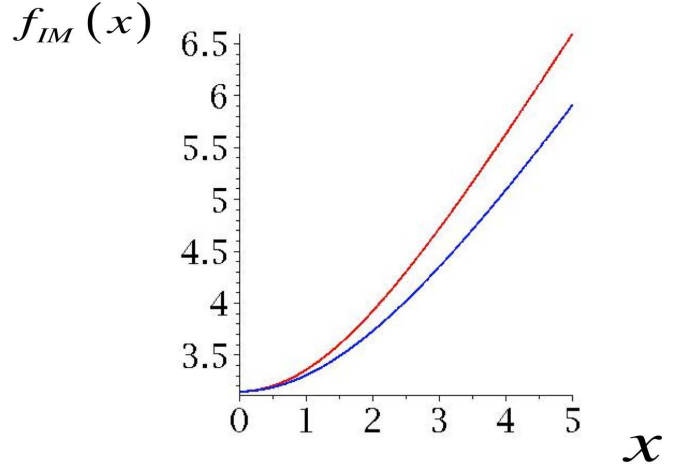}
\caption{(color on-line). Red: exact scaling function $f_{IM}(x)$ in $\I\Sigma^R(\omega,T)$ for the marginal-FL model, Eq.~(\ref{h_3}).
Blue: a scaling function obtained by replacing $\omega\to\sqrt{\omega^2+\pi^2T^2}$ in $\I\Sigma^R(\omega,T=0)$.
Square-root approximation (\ref{h_3_1}) is practically indistinguishable from exact $f_{IM}(x)$ in the interval of $x$ shown in the figure. }
\label{fig:mfl}
\end{figure}

 Expanding
 (\ref{h_3}) in $\omega/T$ and casting the result  into
 the form of a square-root, we obtain, approximately
\be
\mathrm{Im} \Sigma^R_{\bk_F} (\omega,T) \approx \frac{\chi_{L0}}{2\pi v_F} \sqrt{\pi^2 T^2 + \omega^2 \frac{\pi(4-\pi)}{2}}
\label{h_3_1}
\ee
This form is obviously
different  from $\sqrt{\pi^2 T^2 + \omega^2}$
obtained by by replacing $\omega$ by  $\sqrt{\pi^2 T^2 + \omega^2}$ in the $T=0$ result.
Nevertheless, substituting $x = i\pi$ into (\ref{h_3}) we find that $f_{IM} (i\pi)$ vanishes, as it should by the first-Matsubara rule.

The analysis of $\mathrm{Re}\Sigma^R_{\bk_F} (\omega,T) = \omega f_{RM} (\omega/T)$ requires more effort
 as one has to take care
 of
  the
 upper cutoff
 of
  the theory.  The calculation is similar to the one we did for $D=3$ in the previous Section. We use $\I\Sigma^R_{\bk_F} (\omega, T)$
  in the form of Eq.~(\ref{h_2}), but keep the limits of the integration over $\Omega$
  as
  $-E^*$ and $E^*$ and set $E^*$ to infinity only at the end of calculation.
   Without that, we would not reproduce the first-Matsubara rule for  $\mathrm{Re}\Sigma_{\bk_F} (\omega,T)$.
 Substituting $\mathrm{Im}\Sigma^R_{\bk_F}(\omega,T)$
 from Eq.~(\ref{h_2}) into
  the
 KK formula we obtain after some algebra
   \bwt
 \be
   f_{RM} (x) = \frac{2}{\pi} \int_0^\infty \frac{dy}{y^2 -x^2} \left(f_{IM} (y) -y\right) + \frac{2}{\pi}
    \int_0^\infty  \frac{dy}{y^2 -x^2} \left(y - \ln
    \left[1+ e^{y-{\bar E}^*} \right]
    \right),
\label{h_6}
\ee
\ewt
where ${\bar E}^* = E^*/T$.
Both integrals are convergent and are easily evaluated numerically.

At $x = i\pi$, the first integral yields $(2/\pi)
\times
 0.96351$, while the second integral gives $(2/\pi) \left(\ln{{\bar E}^*} + 1 - \ln{\pi}\right)
  \approx
  (2/\pi) \left(\ln{{\bar E}^*} - 0.14473\right)$,
 up to terms exponentially small in $E^*$, which we neglect.
Combining  the two
last expressions, we obtain
\be
f_{RM} (i\pi) \approx \frac{2}{\pi} \left(\ln{{\bar E}^*} + 0.81878\right).
\label{h_7}
\ee

According to the first-Matsubara rule,
 the result in Eq.~(\ref{h_7})
should be exactly the same as
$\R \chi_L ^R(0, T)$
[then $\mathrm{Re} \Sigma^R (i\pi T,T) = i(T/2v_F) f_{RM} (i\pi)$ becomes equal to $i (T/2
v_F) \R\chi_L^R (0,T)$].
 The static local susceptibility is obtained by
 applying the
KK transformation
 to $\I \chi
 ^R
 _L (\omega, T)$
 in Eq.~(\ref{h_4}):
 \bea
 \R
  \chi_L^R (0, T)
  &=& \frac{2}{\pi} \int_0^{{\bar E}^*} \frac{ \tanh{x}}{x} d x 
 \approx \frac{2}{\pi}\left(\ln{\bar E}^*-\int^{\infty}_{0} \frac{dx\ln x}{\cosh^2x}\right)\notag\\
&&\approx  \frac{2}{\pi} \left(\ln {{\bar E}^*} + 0.81878\right),
 \label{h_5}
 \eea
 again, up to terms exponentially small in ${\bar E}^*$.
 Comparing Eqs.~(\ref{h_7}) and (\ref{h_5}), we see that they are equal, as it should be, according to the first-Matsubara rule.

\subsection{
Finite-$q$ instability
}

 The discussion above is valid for a Pomeranchuk instability at $q=0$, when the dynamical exponent $z$ is equal
 to $3$.
 In lattice systems,  an instability may also occur at finite
   $q$, in which case
   $z=2$, up to fluctuation corrections from multi-loop diagrams.\cite{acs,metl_sachdev_2,ac_prl}
   Such an instability is often called
   either
   spin-density-wave (SDW)  or charge-density-wave (CDW), depending on
   whether it occurs in the spin or charge channel.
     The $z=2$ case
    in more involved because typical $q$ along the FS now scale as $q_
    \perp
     \propto |\Omega_n|^{1/2}$, while $q_{
     ||
     }$ still scale as
     $\Sigma (\omega_m + \Omega_n
     ,T
     )$. The one-loop self-energy for $z=2$ problem scales as $\omega^{(D-1)/2}$, hence for $\Omega_n \sim \omega_m$,  typical $q_{
     \perp
     }$ are of order $|\Omega_n|^{(D-1)/2}$.  Local approximation is valid
     if
       typical $q_\parallel
     \ll
      q_{\perp}$, and is only justified for $D >2$. At $D=2$, $q_{\perp}$ and $q_{\parallel}$ are of the same, $|\Omega_n|^{1/2}$, order.
     The local approximation in this case can
     be imposed by extending the system to a large number of fermionic flavors $N$,
      and the analysis up to two loops indeed shows that the local approximation,
       and the first-Matsubara rule associated with it,
         become exact at $N = \infty$.
      For
      a
      $z=2$ transition,
      the
      first-Matsubara rule implies that
      $\Sigma (\omega_m
      ,T
      )$
      evaluated
      at a generic $
      \omega_m \neq \pm \pi T$ contains
      a
      $T^{1/2}$ term (
      or a
      $T^2 \xi^3$
      term
       in the FL regime),
      but the prefactor
      of this term vanishes at  $
      \omega_m
      = \pm \pi T$.  The vanishing is not exact,
    however,
       because
       some of
       the
       higher-order contributions to $\Sigma (\omega_m
       ,T)$  can be viewed as coming from processes  with small momentum transfers, mediated by small $q$ collective excitations of critical $z=2$ modes, and higher-order contributions to $\Sigma (\omega_m,T)$ from
         such processes do not vanish at $N = \infty$ (Refs. \onlinecite{sslee, metl_sachdev,senthil,metl_sachdev_2}).
      Still, in $D=2$, a local propagator  of
      the
       collective mode made of two $z=2$ excitations  scales as
      $\chi_{L}
       (\Omega_n) \propto \int d q d q' d \Omega^{'}_n \chi (q^{'}, \Omega^{'}_n) \chi (q+q', \Omega_n + \Omega^{'}_n) \propto  \ln {|\Omega_n|}$ and is weaker than $\chi_L (\Omega_n) \propto 1/\sqrt{|\Omega_n|}$.   As a result, the prefactor
       of the
        $T^{1/2}$ term in $\Sigma (\omega_m,T)$, although does not vanish exactly at $\omega_m =\pm \pi T$, is nevertheless reduced by
        a factor of
        $|\ln T|/T^{1/2}$.
      Contributions to this prefactor from even higher orders form series in $|\ln T|^n/T^{1/2}$ and may
     potentially give rise to an additional anomalous power
       $T^{\eta}/T^{1/2}$.
       The first-Matsubara rule then remains meaningful as long as $\eta <1/2$.

\section{summary}
\label{sec:concl}

In this paper we analyzed in detail
 the fermionic self-energy
$\Sigma (\omega, T)$  in a FL
at finite temperature $T$ and frequency $\omega$.
 Our
 main
 goal was to understand how general is a certain property of the self-energy,
 the first-Matsubara-frequency rule.
 This rule
 states that the self-energy
  $\Sigma (\omega_m, T)$,
  evaluated at
  discrete
  Matsubara points
   $\omega_m=\pi T(2m+1)$,
    exhibits a special behavior
  at the first fermionic Matsubara frequency
  namely,
 $\Sigma (\pi T, T)$  does not contain terms higher than
 $
 \mathcal{O}
 T$.
 As a particular manifestation of this rule,
  the imaginary part of the self-energy on the FS in a conventional 3D FL
 behaves as
 $\I \Sigma (\omega, T) \propto \omega^2 + \pi^2 T^2$, with exactly
 a
 $\pi^2$ factor
 in front of the
 $T^2$ term, and $\R\Sigma (\omega, T)$ contains
 an
 $\omega$ term but no $\omega T$ term.  We found that the rule is not an exact one, i.e.,
  $\Sigma (\pi T, T)$ in a generic FL does contain
  higher than linear terms in $T$.
  Still, the first term after $\mathcal O(T)$ in $\Sigma (\pi T, T)$ in any dimension $1<D\leq 3$ is of order $T^D$ ($T^3
  \ln T$ in 3D).
    In
    $D >2$, this term is parametrically smaller than $T^2$ term which is present in $\Sigma (\omega_m, T)$
    for
    $|\omega_m|\neq \pi T$.
      We found that the $T^D$ term comes from only forward-
   and  backward scattering, and
     is  expressed in terms of fully renormalized
     amplitudes for these processes.
     We further
     showed
      that the first-Matsubara-frequency rule becomes exact in the
    local approximation,
    when the
       interaction can be approximated by its value
       for the initial and final
        fermionic
        states
        right on the Fermi surface.   In this approximation, which
        is justified,
        e.g.,
         near a Pomeranchuk instability
         even if
      the
      vertex corrections
      are
      non-negligible,  the
   $T^D$ term and all
   higher order
   terms
   in $\Sigma (\pi T, T)$
      vanish, and only
      the
      $O(T)$ term
      survives.
        The first-Matsubara-frequency rule then
    imposes
        two constraints on the
      scaling form of
          the self-energy:
          upon replacing
          $\omega$ by $i\pi T$,
         $\I\Sigma^R (\omega, T)$ must vanish and
            $\R\Sigma^R (\omega, T)$ must reduce to
            an
            $\mathcal{O}(T)$ form.
           We considered several examples
           of the first-Matsubara rule,
           and argued
            that these two
            constraints should be taken into consideration in
             extracting scaling
            forms of
          $\Sigma^R (\omega, T)$ from experimental and numerical data.

\section{acknowledgements}
 Helpful discussions with
D. Basov,
M. Broun,
D. Dessau,
P. Coleman,
 S. Dodge,
 M. Dressel,
 A. Georges,
K. Ingersent,
Y.-B. Kim,
P. Kumar,
 M. Kennett,
 D. van der Marel,
A. Millis,
  U. Nagel, T.
  R{\~o}{\~o}m, M. Sheffler,
 D. Tanner, A.-M. Tremblay, and V. I. Yudson are gratefully acknowledged. The work was supported by NSF-DMR 0906953 and Humboldt foundation (A. V. Ch.), and by NSF-DMR  0908029.
 We are thankful to
 MPIPKS Dresden (A.V. Ch. and D.L.M), the Aspen Center of Physics (A. V. Ch.), the Ruhr-University Bochum (A.V. Ch.),
 Simon Fraser University (D.L.M.), and Swiss NSF
\lq\lq QC2 Visitor Program\rq\rq\/ at
 the University of Basel (D.L.M.)
 for hospitality during the various phases of this work.
 The Aspen Center of Physics is supported in part by the NSF Grant 1066293.

 \appendix
\section{Evaluation of the Matsubara
 self-energy using
 the
 Euler-Maclaurin
 summation formula}
\label{app_a}

In this Appendix, we show how to reproduce the first-Matsubara rule for the fermionic self-energy by using
 the
 Euler-Maclaurin (EM) formula for
 summation over
 the
 Matsubara frequencies.
Unexpectedly, the calculations involving
 the
 EM
 formula turn out to be quite involved, and to reproduce the first-Matsubara rule one has to keep not only the \lq\lq conventional\rq\rq\/ terms in the EM formula, with the integral over a bosonic Matsubara frequency $\Omega_n$ and the  sum over the derivatives of the summand at $n=0$, but also the remainder term, which is often neglected when the EM formula is applied
 in practice.

To be specific,
 we
  consider Eq.~(\ref{m_2}) for the self-energy $\Sigma_{\bk_F} (
 \omega_m,T)$
 and set $\omega_m = \pi T$,
 which gives
 \bea
&&\Sigma_{\bk_F} (\pi T,T) =
i
 T \sum_{
 n
} \int \frac{d q_
 {||}
d^{
 D-1} q_\perp}{(2\pi)^
D} \nonumber \\
 &&\times  \frac{1}{i \pi T (2
 n+1) - v_F q_{||}} \chi_{\bq}(\Omega_n)
\label{1_a}
\eea
with $\bq=(q_{||},\bq_{\perp})$.
 We
  assume that
 the local
 approximation is valid, i.e., that typical $q_{||}$ are small compared
  to typical $q_{\perp}$,
 and
 the dependence
 of the bosonic propagator
 on $q_{||}$
  can be neglected.  Within this approximation, Eq.~(\ref{1_a}) simplifies to
  \bea
&&\Sigma_{\bk_F}
(\pi T,T)
 =
 \frac{T}{2 v_F} \chi_L (0)
\label{12}  \\
+
&& \frac{T}{2 v_F}
 \sum_{n=1}^\infty
 \chi_L
  \left(2\pi T n\right) \left[\mathrm{sgn}(2n+1) - \mathrm{sgn}(2n-1)\right],\nonumber
\eea
where
\be
\chi_L (2\pi T n) = \int d^{D-1} q_\perp \chi
_{\bq_\perp}
 (2\pi n T)/(2\pi)^{D-1}.
\label{14}
\ee
The first term in (\ref{14})
is proportional to
$T$,
and
the second term vanishes identically because for any $n \geq 1$,  $\text{sgn}(2n+1)
= \text{sgn}(2n-1) =1$. Hence,
$\Sigma
_{\bk_F}(\pi T,T)$ does not contain terms beyond $\mathcal{O}(T)$, in accordance with the first-Matsubara rule.

An
unexpected complication arises
 when one attempts to reproduce the vanishing of the
 second term in (\ref{12}) by applying
  the
 EM
 formula
 to
 the sum over $n$.
 Under the condition that ${\bar \chi} (x)$ and its derivatives vanish at $x \to \infty$,
 which we assume to hold
 in our case, the EM formula reads~\cite{kac}
\bea
&&\sum_{n=1}^{\infty} f(n)=\int_{0}^{\infty} f(x)dx - \frac{f(0)}{2}  \nonumber \\
&& - \sum_{p=1}^{N} B_{2p}\frac{f^{(2p-1)}(0)}{(2p)!} - R_N
\label{a_1}
\end{eqnarray}
 where $B_{k}$ are the Bernoulli coefficients, $f^{(n)}$ is the $n$-th
derivative of $f$, and $R_N$ is the
Poisson remainder term
 \be
 R_N = \int_0^\infty \frac{B_N (\{1-x\})}{N!} f^{(N)}(x) dx
\label{a_2}
\ee
where $B_N (x)$ is Bernoulli polynomial, and $\{1-x\}$ denotes the fractional part of $1-x$.

In applications of this formula,
 it is often assumed that  the remainder term $R_N$ tends to zero in the limit $N \to \infty$
 and is thus dropped.
 We show that in our case the remainder term cannot be neglected and one should use the full EM formula, Eqs.~(\ref{a_1} and \ref{a_2}) instead of the truncated one.

Indeed, in our case,
 \be
 f(x) =
 \frac{T}{2 v_F}  \chi_L (2\pi T x) \left[\text{sgn}(2x+1) - \text{sgn}(2x-1)\right].
\label{a_4}
\ee
The first term in the r.h.s. of (\ref{a_1}) is the integral $\int_0^\infty f(x) dx$. Integrating $f(x)$ from (\ref{a_4}) over $x$ we
 obtain
\bea
\int_0^\infty f(x) dx &=&
 \frac{T}{v_F} \int_0^{1/2}
 dx
 \chi_L(2\pi T x) \nonumber \\
 &&=
  \frac{T}{2v_F} \chi_L (0) + \frac{
 \pi T^2}{4
 v_F} \chi'_L (0) + \dots\notag\\
\label{a_5}
\eea
where dots stand for the terms of higher order in $T$.
 Combining (\ref{a_5}) with the boundary term $-f(0)/2=-(T/2v_F)\chi_L(0)$, we
see that the linear-in-$T$ term cancels
 but the quadratic term
contributes
\be
\Sigma_1=
  T^2 \chi'_L(0) \frac{\pi}{4 v_F}
\label{a_6}
\ee
 to
 $\Sigma_{\bk_F}(\pi T,T)$.

This $T^2$ term would violate the first-Matsubara rule and must be canceled
  by the terms with the derivatives $f^{(n)}(0)$.
Because the derivatives of  $\text{sgn}(2x+1)$ and of $\text{sgn}(2x-1)$
 vanish at $x=0$, one has to differentiate only $\chi_L (2\pi T x)$.
A
 $T^2$ contribution to $\Sigma_{\bk_F}(\pi T,T)$ comes from
 the
 first derivative of $f(x)$, i.e., from
 the
  $p=1$ term in the
 sum over $p$ in the r.h.s. of (\ref{a_1}).
 Terms with $p>1$ contribute higher powers of $T$. Using
 that
  $B_2 = 1/6$, we find the $T^2$ contribution from the infinite sum with the derivatives $f^{(n)}(0)$
 as
\be
\Sigma_2 = -
T^2 \chi'_L (0) \frac{\pi}{6 v_F}
\label{a_7}
\ee
The sum $\Sigma_1+\Sigma_2 =
T^2 \chi_L^{'} (0) \pi/(12 v_F)$ is non-zero.

The sum over $p$ in (\ref{a_1}) can be safely extended to infinity as only
the
$p=1$ term contributes
a
 $T^2$ in the self-energy.  If we used the
 truncated
EM formula
 without the remainder term,
 we would
 have
 then obtained
 an
  incorrect result that $\Sigma_{\bk_F} (\pi T,T)$ does contain a $T^2$ term.
 In fact, the
counter-term
 canceling
 the parasitic  $\Sigma_1+\Sigma_2$
 contribution
 does
  come from the remainder term
$R_N$
in (\ref{a_2}), even if we take $N =\infty$ limit.
 Indeed, let's focus on $T^2$ term in the self-energy and replace $\chi_L (2\pi Tx)$ by $2 \pi T \chi'_L (0) x$.
We then have
\be
f(x) \to
\pi \frac{T^2}{2 v_F} x \left[1 - \text{sign} (2x-1)\right].
\label{a_8}
\ee
One can easily make sure that $f(x)$ and its derivatives are non-zero only
in the interval
 $0<x \leq 1/2$,
 where the fractional part of $1-x$ in the argument of the Bernoulli polynomial
 in (\ref{a_2})
 is equal to just $1-x$.
 Furthermore, the derivatives $f^{(n)} (x)$ with $ n \geq 2$ vanish at the boundaries of the integral in (\ref{a_2}),
hence one
can integrating by parts $N-2$ times and
 the boundary terms.
 Using the
  property of Bernoulli polynomials $B'_N (x) = N B_{N-1} (x)$ and applying it $N-2$ times, we rewrite $R_N$ as
  \be
  R_N =
  \frac{T^2}{2v_F} \chi'_L (0) \int_0^1 B_2 (1-
  x)\frac{d^2}{dx^2} \left[
  x (1-\text{sign} (2x-1))\right]
  \label{a_9}
  \ee
 Using that $d/dx\left[1-\text{sgn} (2x-1)\right]
   = -2 \delta (
   x-1/2)$
   and
   also that $B_1 (1/2) =0$
   and $B_2 (1/2) = -1/12$, we obtain after integrating in (\ref{a_9}) by parts
  \be
  \Sigma_3 = -R_N =
  \frac{T^2}{
  v_F} \chi'_L (0)
   B_2 (1/2)
   = -
   T^2 \chi'_L (0) \frac{\pi}{12 v_F}.
  \label{a_10}
  \ee
 Combining the three contributions, we see that $\Sigma_1 + \Sigma_2 + \Sigma_3 =0$, as it should.

 An alternative way to compute the sum over bosonic Matsubara frequencies using
 the
 EM formula would be to
 \lq\lq smear\rq\rq\/
 the discontinuity in $f(x)$ by integrating over $q_{||}$ in (\ref{1_a}) in finite limits
 $-
 Q < q_{||} <
 Q
 $ and take the limit $
 Q\to \infty$ only at the last stage.  In this
  scheme, the remainder term $R_{\infty}$ does not contribute,
  but terms with $p \sim \pi T/Q$ become relevant in the sum over $p$ in (\ref{a_1}).
This calculation is, however, more involved than the one we presented
above, and we
did not find a clear proof that the contribution from $p \sim \pi T/
Q$ exactly cancels $\Sigma_1 + \Sigma_2$.\\

\section{The dependence of the self-energy on the upper cutoff of low-energy theory}
\label{app_new}

In this Appendix we show that the  prefactor of the linear-in-$T$ term in the fermionic self-energy at the first Matsubara frequency
$\Sigma_{\bk_F} (\pi T, T) = \lambda T$
  depends on the ratio of the Fermi energy $E_F = v_F k_F/2$
  to
  the upper cutoff of the low-energy theory
   denoted as  $\Lambda$.
The result
 shown in Eq.~(\ref{ms_2})
 with the \lq\lq mass renormalization factor\rq\rq\/ $\lambda\propto \Pi(0)$
corresponds to
 the situation of
 $\Lambda\ll E_F$, when integration over intermediate energies in the
 expression for the self-energy, Eq. (\ref{u_1}), can be extended to infinity. In the opposite limit of $\Lambda\ll E_F$,
  $\lambda$ is much smaller.
  To see this, we note that typical momentum transfers $
  q =
  |{\bf k}-\bkp|
  $
   are of order $k_F$, hence typical internal energies in the self-energy diagram are of order $E_F$.
   The integration over $\ekp$ in (\ref{u_1}) in finite limits
   changes
   the factor of
   $\mathrm{sgn}(\omega_m+\Omega_n)$ to $(2/\pi) \arctan \left[\Lambda/(\omega_m+\Omega_n)\right]$, which becomes small when
   typical $\Omega_m \sim E_F$ is much larger than $\Lambda$.
   The polarization operator also changes, but the $\Pi (0)$ term
     remains the same
    because
     it comes from
    the smallest frequencies.  To simplify the computations, we keep $\Pi_{\bq} (\Omega_n)$ in the same form as before, but
    replace $v_F q$ by $E_F$, i.e.,
    we
    set
    $\Pi_{\bq} (\Omega_n) =
    -
    \Pi(0)
    (1 - |\Omega_n|/\sqrt{\Omega^2_n + E_F^2})$.
    Substituting this
    expression
    along with the result of
    integration over $\ekp$ into
    the
    self-energy, we obtain
             \bwt
  \bea
&&\Sigma_{\bk_F}(\pi T, T)=
 \lambda T \frac{2}{\pi}
\sum_{\Omega_n} \left(1 - \frac{|\Omega_n|
}{\sqrt{\Omega^2_n + E^2_F}}\right) \arctan{\frac{\Lambda}{\pi T + \Omega_n}} \nonumber \\
&&=
\lambda T \frac{2}{\pi} \left[\arctan{2{\bar \Lambda}} +
\sum^{\infty}_{n=1}\left(1 - \frac{n}{\sqrt{n^2 + {\bar E}_F}}
\right)
 \left
(\arctan{\frac{\bar \Lambda}{n+1/2}} -
\arctan{\frac{\bar \Lambda}{n-1/2}}\right)\right],
\label{ms_3}
\eea
\ewt
where
$\lambda \propto |\Pi_0|$ is the same as in (\ref{ms_2}),
${\bar \Lambda} = \Lambda/ 2\pi T$ and ${\bar E}_F = E_F/2\pi T$.
For ${\bar \Lambda} \gg {\bar E}_F$, the
  term $(1- n/\sqrt{n^2 + {\bar E}^2_F})$
 decreases rapidly for
  $n \gtrsim{\bar E}_F$, when the difference between two
  arctangent
   functions is still small,
   of order
   ${\bar E}_F/{\bar\Lambda}$.
   Then
   the first term in the last line in (\ref{ms_3}) is the dominant one, and using that $ \arctan{2{\bar \Lambda}} \approx \pi/2$ one recovers
  $\Sigma_{\bk_F}(\pi T,T) = \lambda T$
  with $\lambda=\Pi(0)$.
   In the opposite limit
   of
   $ {\bar E}_F\gg {\bar \Lambda}$, the first term can be approximated by
   unity
   for all $n$
    up to $n \sim {\bar E}_F \gg {\bar \Lambda}$. Because the difference
    of
     the two arctangents  scales as $1/n^2$ for $n \gg {\bar \Lambda}$, and
\be
\sum^{\infty}_{n=1} \left(\arctan{\frac{\bar \Lambda}{n+1/2}} -
\arctan{\frac{\bar \Lambda}{n-1/2}}\right) = - \arctan{2{\bar \Lambda}},
\label{la_20}
\ee
the contribution
 to the sum
 from
 positive $n$ almost cancels
 that from $n=0$. A straightforward analysis shows that in this limit, $\lambda$ is small and scales as $\lambda \sim (\Lambda/E_F) \ln{E_F/\Lambda}$.

\section{
Analytic continuation
 of the self-energy
 in the local approximation in 2D}
\label{app_subtle}
 In this appendix we discuss
 some
 subtleties of
   analytic continuation
   of the self-energy
   into
   the
   complex $\omega$ plane in a situation when either $ \mathrm{Re}\Sigma^R(\omega,T)$  or $ \mathrm{Im} \Sigma^R (\omega,T)$ cannot be evaluated explicitly and have to be kept
   in an integral form,
   as in (\ref{41}).
One can use the fact that the integral
  converges in the ultraviolet and modify the integrand by shifting the variable. By doing
  so
  one can obtain several different formulas for$ \mathrm{Im}\Sigma^R(\omega,T)$, which
  all
  nevertheless
  yield the same result along frequency axis.  The danger
  of this trick
  is that, by shifting variables, one imposes the dependence on the external $\omega$ onto $\mathrm{Im}\chi_L^R$ which, in $D \leq 2$, is a non-analytic function of its argument. As a consequence, if one now
  performs analytic continuation
  just by
  replacing $\omega \to z$, one obtains a
 branch cut which stretches down to$ \mathrm{Im} z \to 0$, and the self-energy
  will not obey the first-Matsubara rule at $z = i\pi T$. To
  To make sure that this rule is satisfied,
   one has to use
   the
   Cauchy formula for analytical continuation which,
    in this case,
    is not equivalent to just replacing $\omega$ by a complex $z$.   As an illustration,
    we
     consider
    Eq.~(\ref{41}) in the form
     it was presented in Ref.~\onlinecite{cm_03}
\bwt
 \be
 \mathrm{Im} \Sigma^R(\omega,T) = \frac{B_0}{4} \left( \left(\omega^2 + \pi^2 T^2\right) \ln{\frac{e \Lambda^2}{\pi^2 T^2}}
 + \omega^2\ln{\frac{\pi^2 T^2}{\omega^2}}
+ 1.1217 \pi^2 T^2 + 2 \int_0^\infty \frac{1}{e^{x/T} +1} \left[\omega \ln{\left(\frac{x-\omega}{x+\omega}\right)^2}
  + x \ln{\frac{x^4}{(x^2-\omega^2)^2}}\right]\right).
\label{44}
\ee
\ewt
Along real frequency axis, this formula yields exactly the same
 result as Eq.~(\ref{43}).
However, if we formally replace $\omega$ by $i\pi T + \delta$, with infinitesimally small $\delta >0$,
 before integrating over $x$ in the last term in (\ref{44}), we obtain
$(B_0\pi^2 T^2/4) (0.17 + i \pi)$
 which obviously does not satisfy the first-Matsubara rule.
  The reason is that
  the
  $\omega$ dependence is under the logarithm in the last, integral term of (\ref{44})
, and each of the two logarithms there has
a
branch cut.
Let's
set
$\omega = \pi T e^{i\phi}$ and vary $\phi$ between zero (
the
real axis) and $\pi/2$ (the first Matsubara frequency along
the imaginary axis). To understand what is going on, it is enough to move only little
 off
 the real axis, i.e.,
 to
 consider
 only
 small $\phi$. The branch cuts in the
 first and second
 logarithms in
 the integral term of
Eq.~(\ref{44}) are at $x =1$ and $x = \cos 2 \phi$,
 correspondingly.
 Each
  of them
  gives
  rise to
 a
   discontinuity in the imaginary part
  ($\ln z = \ln |z| + i \pi \mathrm{arg z}$, and
   the
   argument of $z$ changes discontinuously at the branch cut).
    At $\phi =0$,
      the discontinuities coming from the two logarithms cancel
     each other,
     but at
      finite
    $\phi$ there is a range of $x$
    in
    between $1 - \phi^2/4$ and $1$, where the arguments add up to almost $2\pi$
      This additional contribution makes $\mathrm{Im}\Sigma^R( T, \pi T e^{i\phi})$
     to be
      different
      from
      the one obtained by analytical continuation of Eq.~(\ref{44}).
    For  small $\phi$, the difference is $ (
    \pi^2 T^2
    /8) \times (2 \pi
    i) (\phi^2/4)/(e^{\pi} +1)$. We verified numerically that
    this
    expression  is exactly the difference between the analytical continuation of (\ref{44}) and
    the
    brute force replacement $\pi T \to \pi T e^{i\phi}$ in (\ref{44}).

\end{document}